\documentclass[a4paper,11pt]{article}
\pdfoutput=1 % if your are submitting a pdflatex (i.e. if you have
\usepackage[T1]{fontenc} % if needed

\usepackage{graphicx}
\usepackage{xcolor}
\usepackage{tikz}
\usepackage[caption=false]{subfig}
\usepackage{mathrsfs,mathtools}
\usepackage{physics,amssymb}
\usepackage{bm}
\usepackage{braket}
\usepackage{listings}
\usepackage{cases}
\usepackage{comment}
\usepackage{soul}
\usepackage{cancel}
\usepackage{cases}
\usepackage[utf8]{inputenc}
\usepackage{url}
\usepackage{longtable}
\usepackage[normalem]{ulem}
\usepackage{xspace}
\usepackage{jcappub}

\newcommand{\ee}{\mathrm{e}}
\newcommand{\diag}{\mathrm{diag}}

\newcommand{\Mpl}{M_\mathrm{Pl}}
\newcommand{\ns}{n_{\mathrm{s}}}

\newcommand{\SUL}{\mathrm{SU}(2)_\mathrm{L}}

\newcommand{\UY}{\mathrm{U}(1)_\mathrm{Y}}

\newcommand{\sur}{\mathrm{sur}}
\newcommand{\conf}{\mathrm{conf}}
\newcommand{\uend}{\mathrm{end}}

\newcommand{\calH}{\mathcal{H}}

\newcommand{\uJ}{\mathrm{J}}

\newcommand{\calL}{\mathcal{L}}

\newcommand{\calM}{\mathcal{M}}
\newcommand{\calO}{\mathcal{O}}

\newcommand{\calP}{\mathcal{P}}
\newcommand{\calR}{\mathcal{R}}

\newcommand{\beae}[1]{\begin{equation}\begin{aligned} #1 \end{aligned}\end{equation}}

\newcommand{\bae}[1]{\begin{align} #1 \end{align}}
\newcommand{\bce}[1]{\begin{cases} #1 \end{cases}}
\newcommand{\dps}{\displaystyle}

\definecolor{MONZA}{HTML}{CF000F}
\definecolor{DARKBLUE}{HTML}{00008b}
\definecolor{DARKMAGENTA}{HTML}{8b008b}

\hypersetup{colorlinks=true
	,urlcolor=DARKBLUE
	,anchorcolor=DARKBLUE
	,citecolor=DARKBLUE
	,filecolor=DARKBLUE
	,linkcolor=DARKBLUE
	,menucolor=DARKBLUE
	,linktocpage=true
	,pdfproducer=medialab
	,pdfa=true
}

\title{On UV-completion of Palatini-Higgs inflation}

\author[a]{Yusuke Mikura}
\author[a,b]{and Yuichiro Tada}

\affiliation[a]{Department of Physics, Nagoya University, \\
Furocho Chikusaku Nagoya, Aichi 464-8602 Japan}
\affiliation[b]{Institute for Advanced Research, Nagoya University, \\
Furocho Chikusaku Nagoya, Aichi 464-8601 Japan}

\emailAdd{mikura.yusuke@e.mbox.nagoya-u.ac.jp}
\emailAdd{tada.yuichiro@e.mbox.nagoya-u.ac.jp}

\arxivnumber{2110.03925}

\abstract{
We investigate the UV-completion of the Higgs inflation in the metric and the Palatini formalisms. It is known that the cutoff scales for the perturbative unitarity of these inflation models become much smaller than the Planck scale to be consistent with observations. Expecting that the low cutoff scales originate in the curvature of a field-space spanned by the Higgs fields, we consider embedding the curved field-space into a higher dimensional flat space and apply this procedure to the metric-Higgs and the Palatini-Higgs scenarios. The new field introduced in this way successfully flattens the field-space and UV-completes the Higgs inflation in the metric formalism. However, in the Palatini formalism, the new field cannot uplift the cutoff up to the Planck scale. We also discuss the unavoidable low cutoff in the Palatini formalism in the context of the local conformal symmetry. 
}

\begin{document} 
\maketitle
\flushbottom

\section{Introduction}

Cosmic inflation plays a fundamental role in the early universe as it not only solves initial problems of the Big-Bang cosmology but also provides seeds for the large-scale structure of our universe. Among a huge number of inflationary models, the Higgs inflation~\cite{Futamase:1987ua,Bezrukov:2007ep,Barvinsky:2008ia} is one of the most attractive candidates from the viewpoint of minimality because this scenario can be realized within the framework of the Standard Model. In addition to the Standard Model Higgs sector, one often introduces the Higgs-gravity interaction to realize a flat potential which well satisfies the slow-roll condition:
\bae{
\calL\supset \xi(\calH_\uJ^\dag\calH_\uJ)R_\uJ,
}
where $\xi$ is a non-minimal coupling between the Higgs doublet $\calH_\uJ$ and the Ricci scalar $R_\uJ$.

The Ricci scalar in the action is known to be interpreted in two ways: one is called \emph{metric} formalism and the other is referred to as \emph{Palatini} formalism~\cite{Einstein:1925:EFG,Palatini} (see also a recent review~\cite{Tenkanen:2020dge}). These two formulations differ in the definition of the parallel transport. In the metric formalism which is adopted in the usual general relativity, the affine connection is chosen to be metric-compatible and given by the Levi-Civita connection as a function of the metric.
% \Red{In the metric formalism which is adopted in the usual general relativity, the affine connection is chosen to be metric-compatible and symmetric under the change of lower indices (i.e. Torsion-free). These conditions lead to a unique form of the connection called the Levi-Civita connection as a function of the metric.}
Whereas in the Palatini formalism, the connection is rather determined by the Euler--Lagrange constraint of the given action as an auxiliary field.
If the gravity sector is merely given by the Einstein--Hilbert action, two formulations become equivalent because the Euler--Lagrange constraint leads to the Levi-Civita connection. However, if the non-minimal coupling is present as in the Higgs inflation, two formulations result in different physical consequences even though the apparent expression of the action does not change in the Jordan frame. 
To distinguish these two formulations in the Higgs inflation, we call the original Higgs inflation in the metric formalism \emph{metric-Higgs inflation} and the Palatini variant \emph{Palatini-Higgs inflation}~\cite{Bauer:2008zj}.

Phenomenology in both the metric-Higgs and the Palatini-Higgs inflation has been well investigated so far and inflationary predictions in both scenarios are known to be in perfect agreement with observations on the cosmic microwave background (CMB) conducted by the Planck Collaboration~\cite{Planck:2018jri}.\footnote{For recent reviews, see Ref.~\cite{Rubio:2018ogq} for the metric-Higgs and Ref.~\cite{Tenkanen:2020dge} for the Palatini-Higgs inflation.} The CMB observations however require that the coupling to gravity $\xi$ should be unnaturally large $\xi\gg1$ if the self-coupling of the Higgs doublet $\lambda$ is not tuned to be quite small. Such a large $\xi$ induces a lower cutoff $\Lambda$ compared to the Planck scale and leads to discussions on a violation of the perturbative unitarity~\cite{Burgess:2009ea,Barbon:2009ya,Barvinsky:2009ii,Burgess:2010zq,Hertzberg:2010dc,Bezrukov:2010jz,Bauer:2010jg,Lerner:2011it,Kehagias:2013mya,Ren:2014sya,McDonald:2020lpz,Hamada:2020kuy,Antoniadis:2021axu}. 
In the metric-Higgs scenario, the cutoff during inflation $\Lambda\sim\Mpl/\sqrt{\xi}$ is sufficiently larger than the Hubble scale $H\sim\Mpl/\xi$ as we will review and the inflationary dynamics is expected to be predictable~\cite{Bezrukov:2010jz},
while in the reheating phase it gets lower as $\Lambda\sim\Mpl/\xi$ but the relevant energy scale becomes larger as $\sim\Mpl/\sqrt{\xi}$ which means that the theory must be corrected in order to precisely predict the reheating dynamics~\cite{Ema:2016dny}.
In the Palatini-Higgs model, the cutoff is universally given by $\Lambda\sim\Mpl/\sqrt{\xi}$ and thus it would be predictable over the whole phases of the early universe.

As it requires a correction in the reheating phase, the UV-completion of the metric-Higgs inflation is practically an urgent task. Ref.~\cite{Giudice:2010ka} showed that the low cutoff in the metric-Higgs model is caused by the large field-space curvature of the Higgs and an additional scalar which flattens their field-space can uplift the cutoff scale up to the Planck scale.
Such a scalar can be understood as the scalaron in the gravity sector made dynamical by the additional $R^2$ term in the Lagrangian, which is necessary to one-loop renormalizability~\cite{Salvio:2015kka,Ema:2017rqn,Gorbunov:2018llf,Ema:2020zvg}. Accordingly, the metric-Higgs-$R^2$ system is applicable up to the Planck scale and now the corresponding inflation is fully predictable.

Similarly, even though it is not practically problematic, the UV-completion of the Palatini-Higgs inflation is of theoretical interest to investigate whether it is a natural low-scale effective field theory (EFT).\footnote{
Ref.~\cite{Jinno:2019und} pointed out that the Palatini-Higgs inflation is sensitive to higher dimensional operators, so that it is challenging to realize as a low-energy EFT.
}
In this paper, we study the possibility of the weakly coupled UV-completion of the Palatini-Higgs inflation. We first provide a straightforward approach to embed the curved field-space into a higher dimensional flat space in the Einstein frame without specifying gravitational formulations. 
We then show that the added field successfully UV-completes the metric-Higgs inflation, but it hardly uplifts the cutoff scale in the Palatini-Higgs inflation. Thus we would conclude the low cutoff in the Palatini-Higgs model does not simply originate from the field-space curvature.

This paper is organized as follows. We first define our setup in Sec.~\ref{setup}. We then review the phenomenology of the metric-Higgs and Palatini-Higgs inflation in Sec.~\ref{inf. pheno.}. The unitarity issue caused by the large non-minimal coupling is summarized in Sec.~\ref{perturbative unitarity}. We discuss the UV-completion in Sec.~\ref{sec3}. In particular, the embedding approach and its applications to the metric-Higgs and the Palatini-Higgs scenarios are studied in Sec.~\ref{sec3-1}. The unavoidable low cutoff is discussed in terms of the local conformal symmetry in Sec.~\ref{sec3-2}. The conclusions are presented in Sec.~\ref{sec4}.

\section{Higgs inflation in metric and Palatini}\label{sec2}

\subsection{Setup}\label{setup}

The theory of interest is composed of the Standard Model Higgs doublet $\calH_\uJ$ with the non-minimal coupling to gravity:\footnote{Throughout this paper, we adopt the natural unit $c=\hbar=1$ and $\eta_{\mu\nu}=\diag(-1,1,1,1)$ is used as the sign of the Minkowski metric.}
\bae{\label{starting action}
    S=\int\dd[4]{x}\sqrt{-g_\uJ}\left[\frac{\Mpl^2}{2}\left(1+2\xi\frac{\calH_\uJ^\dag\calH_\uJ}{\Mpl^2}\right)R_\uJ+\calL_\mathrm{SM}\right],
}
where $\Mpl$ is the reduced Planck mass, $\xi$ is a coupling constant, $R_\uJ$ is the Ricci scalar, and $\calL_\mathrm{SM}$ denotes the Standard Model Lagrangian.\footnote{Throughout this paper, all quantities with the subscript $\uJ$ are defined in the Jordan frame.} For the Standard Model Lagrangian, we focus on the gauge-Higgs sector in this paper whose Lagrangian is given by
\bae{\label{gauge-Higgs}
    \calL_\mathrm{SM}\supset-\frac{1}{4}A_{\mu\nu}^aA^{a\mu\nu }-\frac{1}{4}B_{\mu\nu}B^{\mu\nu}-g_\uJ^{\mu\nu}(D_\mu \calH_\uJ)^\dag (D_\nu\calH_\uJ)-\lambda(\calH_\uJ^\dagger\calH_\uJ)^2,
}
where $\lambda$ denotes the self-coupling of the Higgs doublet, $A^a_\mu$ is a set of three $\SUL$ gauge bosons $(a=1,2,3)$ with their field strengths $A^a_{\mu\nu}$, and $B_\mu$ is the $\UY$ gauge boson with its field strength $B_{\mu\nu}$. The field strengths of $\SUL$ and $\UY$ gauge bosons are respectively expressed with structure constants $\epsilon_{abc}$ as
\bae{
A^a_{\mu\nu}=\partial_\mu A^a_\nu-\partial_\nu A^a_\mu+g\epsilon_{abc}A^b_\mu A^c_\nu \qc
    B_{\mu\nu}=\partial_\mu B_\nu-\partial_\nu B_\mu.
}
Here the gauge covariant derivative is defined by
\bae{
D_\mu \coloneqq \pqty{\partial_\mu-ig\frac{\tau^a}{2}A^a_\mu-\frac{i}{2}g^\prime B_\mu},
}
where $g$ and $g^\prime$ are the $\SUL$ and $\UY$ gauge couplings and $\tau^a$ are the Pauli matrices defined by
\beae{
    \tau^1=\pmqty{0 & 1 \\ 1 & 0} \qc
    \tau^2=\pmqty{0 & -i \\ i & 0} \qc
    \tau^3=\pmqty{1 & 0 \\ 0 & -1}.
}
The mass term of the Higgs doublet is neglected throughout this paper because it is irrelevant during inflation and reheating.

As mentioned in the introduction, though the metric formalism is adopted as a standard gravitational formulation, one can also interpret gravity by the Palatini approach instead which leads to different phenomenological consequences in a theory with a non-minimal coupling to gravity. The differences are explicit and easily understandable in the so-called Einstein frame where the non-minimal coupling is removed from the theory by a conformal redefinition of the metric
\bae{\label{Eq. conformal redefinition}
   g_{\mu\nu}\coloneqq \Omega^2 g_{\uJ\mu\nu},\qquad \Omega^2\coloneqq \pqty{1+2\xi\frac{\calH_\uJ^\dag\calH_\uJ}{\Mpl^2}}.
}
Accompanied by the metric redefinition, the connection also transforms in the metric formalism because it is given by the Levi-Civita connection:
\bae{
 \Gamma^\rho_{\mu\nu}(g)=\frac{1}{2}g^{\rho\lambda}\pqty{\partial_\mu g_{\nu\lambda}+\partial_\nu g_{\mu\lambda}-\partial_\lambda g_{\mu\nu}}.
 }
Meanwhile, the connection is left unaffected in the Palatini formalism since it is treated as an independent variable as well as the metric. Thus, the Ricci scalar transforms differently depending on the underlying gravitational formulations as~\cite{Tenkanen:2020dge}
\bae{
    \sqrt{-g_\uJ}\Omega^2R_\uJ=\sqrt{-g}(R-6\kappa\Omega g^{\mu\nu}\nabla_\mu\nabla_\nu \Omega^{-1}),
}
where $\kappa=1$ and $\kappa=0$ correspond to the metric and the Palatini formalism, respectively. After the rescaling of the metric, we get the Einstein frame expression:
\bae{\label{full einstein}
    S=\int\dd[4]{x}\sqrt{-g}\left[\frac{\Mpl^2}{2}R-3\kappa\Mpl^2\Omega g^{\mu\nu}\nabla_\mu\nabla_\nu\Omega^{-1}-\frac{1}{4}A_{\mu\nu}^aA^{a\mu\nu }-\frac{1}{4}B_{\mu\nu}B^{\mu\nu}\right. \nonumber \\
    \left.-\frac{1}{\Omega^2}g^{\mu\nu}(D_\mu \calH_\uJ)^\dag (D_\nu\calH_\uJ)-\frac{\lambda}{\Omega^4}(\calH_\uJ^\dagger\calH_\uJ)^2\right],
}
where the field strengths of the $\SUL$ and $\UY$ gauge fields are left unchanged thanks to their conformal invariance. In the Einstein frame, the connection does not directly couple to the Higgs field $\calH_\uJ$ and the gravity sector is merely the Einstein--Hilbert form. In such a case, the Euler--Lagrange constraint in the Palatini formalism restricts the connection to the Levi-Civita one, and the two formulations become equivalent, up to the explicit difference in the $\kappa$ term~\cite{Tenkanen:2020dge}.\footnote{We assume the torsion-free condition throughout this paper.}

\subsection{Inflationary phenomenology}\label{inf. pheno.}

Let us first review phenomenological aspects of the metric-Higgs inflation~\cite{Bezrukov:2007ep} and the Palatini-Higgs inflation~\cite{Bauer:2008zj,Takahashi:2018brt}. In this subsection, we neglect the gauge sector for simplicity, which is justified later.

In the inflationary literature, we usually take the \emph{unitary gauge} in which the Higgs doublet is described by a real scalar field $\phi_\uJ(x)$ as $\calH_\uJ^T(x)=(0,\phi_\uJ(x)/\sqrt{2})$. With use of the unitary gauge, the Higgs sector in Eq.~\eqref{full einstein} is reduced to
\bae{\label{E-frame unitary}
    S_\mathrm{inf}=\int\dd[4]{x}\sqrt{-g}\bqty{\frac{\Mpl^2}{2}R-\frac{1+\xi\frac{\phi_\uJ^2}{\Mpl^2}+6\kappa\xi^2\frac{\phi_\uJ^2}{\Mpl^2}}{2\left(1+\xi\frac{\phi_\uJ^2}{\Mpl^2}\right)^2}g^{\mu\nu}\partial_\mu\phi_\uJ\partial_\nu\phi_\uJ-\frac{\lambda \phi_\uJ^4}{4\left(1+\xi\frac{\phi_\uJ^2}{\Mpl^2}\right)^2}},
}
where $\kappa=1$ for the metric-Higgs and $\kappa=0$ for the Palatini-Higgs inflation. The non-trivial kinetic term can be canonically normalized by introducing the field $\chi$ defined through
\bae{\label{canonical transformation}
    \dv{\phi_\uJ}{\chi}\coloneqq \sqrt{\frac{\left(1+\xi\frac{\phi_\uJ^2}{\Mpl^2}\right)^2}{1+\xi\frac{\phi_\uJ^2}{\Mpl^2}+6\kappa\xi^2\frac{\phi_\uJ^2}{\Mpl^2}}}.
}
In terms of $\chi$, the action can be rewritten as 
\bae{
S_\mathrm{inf}=\int\dd[4]{x}\sqrt{-g}\bqty{\frac{\Mpl^2}{2}R-\frac{1}{2}g^{\mu\nu}\partial_\mu\chi\partial_\nu\chi-U(\chi(\phi_\uJ))},
}
with the potential in the Einstein frame
\bae{
U(\chi(\phi_\uJ))\coloneqq\frac{\lambda \phi_\uJ^4(\chi)}{4\left(1+\xi\frac{\phi_\uJ^2(\chi)}{\Mpl^2}\right)^2}.
}
The change of variable~\eqref{canonical transformation} can be easily integrated in the Palatini case, while an asymptotic form in the large field limit $\xi\phi_\uJ^2/\Mpl^2\gg1$ is useful in the metric case as
\bae{
\mathrm{metric}&:\quad\phi_\uJ\simeq \frac{\Mpl}{\sqrt{\xi}}\mathrm{exp}\left(\sqrt{\frac{1}{6}}\frac{\chi}{\Mpl}\right),\\
\mathrm{Palatini}&:\quad \phi_\uJ=\frac{\Mpl}{\sqrt{\xi}}\mathrm{sinh}\left(\frac{\sqrt{\xi}\chi}{\Mpl}\right).
}
The potential is reduced to
\bae{\label{metric potential}
    \mathrm{metric}&:\quad U\simeq \frac{\lambda\Mpl^4}{4\xi^2}\left(1+\mathrm{exp}\left(-\sqrt{\frac{2}{3}}\frac{\chi}{\Mpl}\right)\right)^{-2}, \\
    \label{Palatini potential}
    \mathrm{Palatini}&:\quad U=\frac{\lambda\Mpl^4}{4\xi^2}\mathrm{tanh}^4\left(\frac{\sqrt{\xi}\chi}{\Mpl}\right).
}
The potentials in both scenarios approach asymptotically to a constant value $U\simeq \frac{\lambda\Mpl^4}{4\xi^2}$ at a large field region, which is suitable for slow-roll inflation.

Let us compute inflationary observables of the Higgs inflation. The inflaton's dynamics is characterized by slow-roll parameters and the backward e-folds $N$ defined by the canonical field as
\bae{
    \epsilon\coloneqq\frac{\Mpl^2}{2}\left(\frac{1}{U}\dv{U}{\chi}\right)^2,\qquad \eta\coloneqq\Mpl^2\left(\frac{1}{U}\dv[2]{U}{\chi}\right), \qquad N(\chi)\coloneqq \frac{1}{\Mpl^2}\int^\chi_{\chi_\mathrm{end}} U\left(\dv{U}{\chi}\right)^{-1}\dd{\chi},
}
where $\chi_\mathrm{end}$ denotes the field value of $\chi$ at the end of inflation. The primary cosmological observables, the spectral index $\ns$ and the tensor-to-scalar ratio $r$, are expressed to the leading order in the slow-roll parameters by
\bae{
\ns\simeq 1-6\epsilon+2\eta, \qquad r\simeq 16\epsilon.
}
In terms of the number of e-folds, the resulting predictions to the leading order in large $N$ are given by 
\bae{
\mathrm{metric}&:\quad (\ns,r)\simeq \left(1-\frac{2}{N},\frac{12}{N^2}\right),\\
\mathrm{Palatini}&:\quad (\ns,r)\simeq \left(1-\frac{2}{N},\frac{2}{\xi N^2}\right).
}
With the typical e-folds $N\sim50\text{--}60$, the predicted spectral index in both scenarios is well consistent with CMB observations~\cite{Planck:2018jri}.
The biggest phenomenological difference between the two scenarios lies in the tensor-to-scalar ratio. In the metric-Higgs scenario, without depending on the coupling constant $\xi$, it converges to a constant value of order $10^{-3}$ which is within the reach of the future observations~\cite{Hazumi:2019lys,CMB-S4:2016ple,CMB-S4:2020lpa}. Whereas the value of the tensor-to-scalar ratio in the Palatini-Higgs scenario can be extremely small due to suppression in the large $\xi$ limit.

The value of the coupling constant $\xi$ cannot be arbitrarily chosen due to the observational constraint on the dimensionless power spectrum of curvature perturbation $\calP_\zeta$~\cite{Lyth:1998xn,Planck:2018vyg} defined by
\bae{
\calP_\zeta= \frac{1}{24\pi^2\Mpl^4}\frac{U}{\epsilon},
}
where $U$ is the potential energy and $\epsilon$ is the slow-roll parameter. An observed amplitude $\calP_\zeta\simeq 2.1\times 10^{-9}$~\cite{Planck:2018jri} fixes the relation between $\xi$ and $\lambda$ as
\bae{\label{metric xi}
\mathrm{metric}&:\quad \xi\simeq \sqrt{\frac{\lambda}{72\pi^2\calP_\zeta}}N\sim 5\times10^4\sqrt{\lambda},\\
\mathrm{Palatini}&:\quad \xi\simeq \frac{\lambda N^2}{12\pi^2\calP_\zeta}\sim10^{10}\lambda.
}
The CMB normalization restricts that the coupling to gravity $\xi$ should be quite large unless the quartic coupling $\lambda$ is extremely small both in the metric and Palatini formalisms.\footnote{There exists a possibility that the self-coupling of the Higgs field $\lambda$ is tuned to be small at inflationary energy scales, which is known as the critical Higgs scenario~\cite{Hamada:2014iga,Bezrukov:2014bra,Hamada:2014wna,Enckell:2020lvn}.}

\subsection{Perturbative unitarity}\label{perturbative unitarity}

As seen in the previous subsection, the coupling to gravity $\xi$ needs to be quite large to realize an observed amplitude of the curvature perturbation both in the metric-Higgs and Palatini-Higgs scenarios. The large value of $\xi$, however, can be problematic from a viewpoint of theoretical consistency because it may cause the unitarity violation which is interpreted as a breakdown of the validity of the conventional perturbative analysis.

In this subsection, we will encapsulate the unitarity issue of the metric-Higgs inflation~\cite{Burgess:2009ea,Barbon:2009ya,Barvinsky:2009ii,Burgess:2010zq,Hertzberg:2010dc,Bezrukov:2010jz,Lerner:2011it,Kehagias:2013mya,Ren:2014sya,Hamada:2020kuy} and the Palatini-Higgs inflation~\cite{Bauer:2010jg,McDonald:2020lpz} in two ways of gauge fixing. Although one may study in any frame because the unitarity violation should be a frame-independent phenomenon, in what follows, we insist on using the Einstein frame.

\subsubsection{Unitary gauge}\label{unitary gauge}
We first study how the cutoff appears if one takes the unitary gauge. In general, the Higgs's kinetic term, the gauge-Higgs interaction, and the Higgs's quartic potential can serve as distinct sources of the unitarity violation.\footnote{For completeness, we must add the fermion sector.} However, with use of the unitary gauge, the non-trivial kinetic term in Eq.~\eqref{E-frame unitary} can be canonically normalized by the field redefinition as mentioned in Sec.~\ref{inf. pheno.}. Thus, it is enough to examine the gauge-Higgs interaction and the Higgs potential written in the canonical field $\chi$ to estimate the cutoff scale. Associated terms before defining a new canonical field are given by
\bae{\label{matter Lagrangian by Higgs}
\calL\supset -\frac{g^2}{8}\frac{\phi_\uJ^2}{\Omega^2}\left[(A^1_\mu)^2+(A^2_\mu)^2+\frac{1}{\cos^2\theta}(Z_\mu)^2\right]-\frac{\lambda}{4\Omega^4}\phi_\uJ^4,
}
where we define
\bae{
\tan \theta\coloneqq \frac{g^\prime}{g},\qquad  Z_\mu\coloneqq \cos\theta A_\mu^3-\sin\theta B_\mu.
}

During inflation $\xi\phi_\uJ^2\gg\Mpl^2$,
since $\Omega^2(\phi_\uJ)$ can be well approximated as $\Omega^2\simeq \xi\phi_\uJ^2/\Mpl^2$, the gauge-Higgs interaction and the Higgs potential term in the Lagrangian~\eqref{matter Lagrangian by Higgs} are respectively reduced to mass terms of the gauge fields and the cosmological constant as
\bae{
    \calL\supset -\frac{g^2\Mpl^2}{8\xi}\left[(A^1_\mu)^2+(A^2_\mu)^2+\frac{1}{\cos^2\theta}(Z_\mu)^2\right]-\frac{\lambda\Mpl^4}{4\xi^2}.
}
We expect that the scattering processes with the longitudinal modes break the perturbative unitarity once the energy of our interest becomes greater than their mass scale $\Mpl/\sqrt{\xi}$~\cite{Bezrukov:2010jz}. Therefore, the mass scale of the gauge fields can be understood as a cutoff of the theory $\Lambda$ during inflation:
\bae{\label{eq: cutoff during inflation}
    \eval{\Lambda}_{\xi\phi_\uJ^2/\Mpl^2\gg 1}\sim \frac{\Mpl}{\sqrt{\xi}}.
}
Note that the scale $\Lambda$ serves as a cutoff regardless of the choice of gravitational formulations because all terms related to the gauge fields are unaffected by the scalar's field redefinition.

% We turn to estimate the violation scale at the reheating epoch where the Higgs field $\phi$ oscillates about its origin. Expanding the gauge-Higgs interaction and the quartic potential with respect to $\chi$ around $\chi=0$ and seeing the higher-dimensional operators, one obtains the cutoff at the small field regime as
% \bae{
%     \mathrm{metric}&:\quad \eval{\Lambda}_{\phi_\uJ^2\to0}\sim \frac{\Mpl}{\xi}, \label{eq: cutoff around origin in metric} \\
%     \mathrm{Palatini}&:\quad \eval{\Lambda}_{\phi_\uJ^2\to0}\sim \frac{\Mpl}{\sqrt{\xi}},
% }
% which are much smaller than the Planck scale $\Mpl$ in the large $\xi$. 

We turn to estimate the violation scale at the reheating epoch where the Higgs field $\phi_\uJ$ oscillates about its origin. To this end, let us expand the gauge-Higgs interaction and the quartic potential with respect to the canonical field $\chi$ around $\chi=0$. With use of Eq.~\eqref{canonical transformation}, the Higgs field $\phi_\uJ$ can be expressed by series as
\bae{
    \mathrm{metric}&:\quad \phi_\uJ\simeq \chi-\xi^2\frac{\chi^3}{\Mpl^2}+\cdots, \\
    \mathrm{Palatini}&:\quad \phi_\uJ\simeq \chi+\frac{1}{6}\xi\frac{\chi^3}{\Mpl^2}+\cdots,
}
which give rise to
\bae{
    \mathrm{metric}:&\quad -\frac{g^2}{8}\frac{\phi_\uJ^2}{\Omega^2}A^2-\frac{\lambda}{4}\frac{\phi_\uJ^4}{\Omega^4}\simeq -\frac{g^2}{8}\chi^2 A^2-\frac{\lambda}{4}\chi^4+\frac{g^2}{4}\frac{\chi^4 A^2}{(\Mpl/\xi)^2}+\lambda\frac{\chi^6}{(\Mpl/\xi)^2}+\cdots, \nonumber \\
    \mathrm{Palatini}:&\quad-\frac{g^2}{8}\frac{\phi_\uJ^2}{\Omega^2}A^2-\frac{\lambda}{4}\frac{\phi_\uJ^4}{\Omega^4}\simeq -\frac{g^2}{8}\chi^2 A^2-\frac{\lambda}{4}\chi^4+\frac{g^2}{12}\frac{\chi^4 A^2}{(\Mpl/\sqrt{\xi})^2}+\frac{\lambda}{3}\frac{\chi^6}{(\Mpl/\sqrt{\xi})^2}+\cdots,
}
where $A^2$ collectively denotes $(A^1_\mu)^2+(A^2_\mu)^2+\frac{1}{\cos^2\theta}(Z_\mu)^2$. One can see that higher dimensional operators are suppressed by $\Mpl/\xi$ in the metric-Higgs scenario while the scale in denominators in the Palatini-Higgs one is $\Mpl/\sqrt{\xi}$. So the cutoff scales at the small field regime can be respectively read off as
\bae{
    \mathrm{metric}&:\quad \eval{\Lambda}_{\phi_\uJ^2\to0}\sim \frac{\Mpl}{\xi}, \label{eq: cutoff around origin in metric} \\
    \mathrm{Palatini}&:\quad \eval{\Lambda}_{\phi_\uJ^2\to0}\sim \frac{\Mpl}{\sqrt{\xi}},
}
which are much smaller than the Planck scale $\Mpl$ in the large $\xi$.

% \bae{
% -\frac{g^2}{8}\frac{\phi_\uJ^2}{\Omega^2}A^2-\frac{\lambda}{4}\frac{\phi_\uJ^4}{\Omega^4}\simeq -\frac{g^2}{8}\chi^2 A^2-\frac{\lambda}{4}\chi^4+\frac{g^2}{12}\frac{\chi^4 A^2}{(\Mpl/\sqrt{\xi})^2}+\frac{\lambda}{3}\frac{\chi^6}{(\Mpl/\sqrt{\xi})^2}\nonumber
% \\
% -\frac{17 g^2}{360}\frac{\chi^6 A^2}{(\Mpl/\sqrt{\xi})^4}-\frac{3\lambda}{10}\frac{\chi^8}{(\Mpl/\sqrt{\xi})^4}
% }

\subsubsection{Covariant gauge}\label{sec: covariant gauge}

Instead of the unitary gauge, one can parametrize the Higgs doublet $\calH_\uJ$ in terms of four real scalar fields as
\bae{
    \calH_\uJ=\frac{1}{\sqrt{2}}\pmqty{\phi_{\uJ 1}+i\phi_{\uJ 2} \\
    \phi_{\uJ 3}+i\phi_{\uJ 4}},
}
which we call the \emph{covariant gauge}~\cite{Burgess:2010zq}. In the following, we will study how the cutoff scale is determined with this parametrization. To this end, we start with the Einstein frame action~\eqref{full einstein}. By writing the gauge covariant derivative explicitly, the action becomes
\bae{\label{Eq. explicit covariant}
    S=\int\dd[4]{x}\sqrt{-g}\left[\frac{\Mpl^2}{2}R-\frac{1}{4}A_{\mu\nu}^aA^{a\mu\nu}-\frac{1}{4}B_{\mu\nu}B^{\mu\nu}-\frac{1}{2}G_{ij}(\phi_\uJ)\partial_\mu\phi_{\uJ i}\partial^\mu\phi_{\uJ j}-\frac{\lambda}{\Omega^4}(\calH_\uJ^\dagger\calH_\uJ)^2 \right. \nonumber \\
    \left.+\frac{1}{\Omega^2}\left\{A^a_\mu J^{a\mu}_{\uJ A}+B_\mu J^\mu_{\uJ B}+\frac{1}{2}g g^\prime B^\mu A_\mu^a \calH_\uJ^\dagger \tau^a \calH_\uJ-\frac{1}{4}{g^\prime}^2\calH_\uJ^\dagger\calH_\uJ(B_\mu)^2-\frac{1}{4}g^2\calH_\uJ^\dagger\calH_\uJ(A_\mu^a)^2\right\}\right],
}
where $J_{\uJ A}^{a\mu}$ and $J_{\uJ B}^\mu$ are currents defined by
\bae{
    J_{\uJ A}^{a\mu}\coloneqq i\frac{g}{2}(\calH_\uJ^\dagger\tau^a\partial^\mu\calH_\uJ-\partial^\mu \calH_\uJ^\dagger \tau^a \calH_\uJ) \qc J_{\uJ B}^\mu\coloneqq i\frac{g^\prime}{2}(\calH_\uJ^\dagger\partial^\mu\calH_\uJ-\partial^\mu \calH_\uJ^\dagger \calH_\uJ),
}
and $G_{ij}(\phi_\uJ)$ is the field-space metric on the curved field-space $\calM_4$. An explicit form of the field-space metric is given by
\bae{\label{field-space}
    G_{ij}(\phi_\uJ)=\frac{1}{1+\xi\frac{\phi_{\uJ k}^2}{\Mpl^2}}\pqty{\delta_{ij}+\frac{6\kappa\xi^2\frac{\phi_{\uJ i}\phi_{\uJ j}}{\Mpl^2}}{1+\xi\frac{\phi_{\uJ k}^2}{\Mpl^2}}}.
}

When the covariant gauge is adopted, the Higgs kinetic term can not be diagonalized at all once as opposed to the case of the unitary gauge due to the non-vanishing curvature invariant of the field-space metric as we see below. Thus, the Higgs kinetic term, the Higgs potential, and the gauge-Higgs interaction can all be sources of the unitarity violation. In the covariant gauge, we instead consider removing the higher-dimensional operators from the Higgs potential and the gauge-Higgs interaction. Indeed, this can be realized by defining a new doublet $\calH$ via $\calH\coloneqq \calH_\uJ/\Omega$. In terms of $\calH$, one can see that the Higgs potential and the gauge-Higgs interaction are all written by four-dimensional operators, making them irrelevant in the discussion on the unitarity issue once $\calH$ is used.\footnote{Note particularly that the currents transform conformal-covariantly as
\bae{
    J_A^{a\mu}=i\frac{g}{2}(\calH^\dagger\tau^a\partial^\mu\calH-\partial^\mu\calH^\dagger\tau^a\calH)=\frac{1}{\Omega^2}J_{\uJ A}^{a\mu} \qc J_B^\mu=\frac{1}{\Omega^2}J_{\uJ B}^\mu.
}}
Thus, one can determine the cutoff of the theory by studying only the Higgs kinetic term.

Now that the field-space $\calM_4$ is curved, the curvature $\calR_G$ of the field-space naturally implies the cutoff scale~\cite{DeCross:2016cbs,Sfakianakis:2018lzf}. Thus, the violation scale $\Lambda(\phi_\uJ)$ can be read out via $\Lambda\sim \calR_G^{-1/2}$. Indeed it has been confirmed that the cutoff associated with the scattering amplitude is consistent with the field-space curvature interpretation in the small field region (see, e.g., Refs.~\cite{Alonso:2015fsp,Nagai:2019tgi}).
It should be noted that, since the curvature is invariant under the coordinate transformation, one can safely calculate the curvature with $\calH_\uJ$ instead of $\calH$. Given the field-space metric~\eqref{field-space}, the curvature $\calR_G$ can be calculated as
\bae{\label{metric target curvature}
	\mathrm{metric}&:\quad \calR_G=\frac{6\xi\left\{4+12\xi+(5\xi+36\xi^2+36\xi^3)\frac{\phi_{\uJ i}^2}{\Mpl^2}+\xi^2(1+6\xi)^2\frac{\phi_{\uJ i}^4}{\Mpl^4}\right\}}{\Mpl^2\left\{1+\xi(1+6\xi)\frac{\phi_{\uJ i}^2}{\Mpl^2}\right\}^2},\\
	\label{palatini target curvature}\mathrm{Palatini}&:\quad \calR_G=\frac{6\xi\left(4+\xi\frac{\phi_{\uJ i}^2}{\Mpl^2}\right)}{\Mpl^2\left(1+\xi\frac{\phi_{\uJ i}^2}{\Mpl^2}\right)},
}
and the cutoff scales for the perturbative unitairity read
\bae{\label{cutoff metric}
	\mathrm{metic}:&\quad \eval{\Lambda}_{\phi_{\uJ i}^2\to0}\sim \frac{\Mpl}{\xi}, \qquad \eval{\Lambda}_{\xi\phi_{\uJ i}^2/\Mpl^2\gg1}\sim \frac{\Mpl}{\sqrt{\xi}}, \\
	\label{cutoff Palatini}\mathrm{Palatini}:&\quad \eval{\Lambda}_{\phi_{\uJ i}^2\to0}\sim \frac{\Mpl}{\sqrt{\xi}}, \qquad \eval{\Lambda}_{\xi\phi_{\uJ i}^2/\Mpl^2\gg1}\sim \frac{\Mpl}{\sqrt{\xi}},
}
showing the same results derived by taking the unitary gauge.

\subsubsection{Unitarity issue during inflation and reheating}

We have seen that the cutoff of the theory becomes much smaller than the Planck scale due to the large coupling to gravity. Here let us study if the unitarity violation really occurs during inflation or at the reheating stage. 

We first address the issue during inflation. Particles are excited mainly by the gravitational interaction and thus their energies are typically characterized by the Hubble scale $H$ in this case. One can hence say that the perturbative unitarity is violated if the Hubble scale exceeds the cutoff $\Lambda$ evaluated at a large field value. In both metric-Higgs and Palatini-Higgs scenarios, the potential energy $U$ is of order $\lambda\Mpl^4/\xi^2$ and, correspondingly, the value of $H\sim U^{1/2}/\Mpl$ is approximated as $\sqrt{\lambda}\Mpl/\xi$ (see Eqs.~\eqref{metric potential} and~\eqref{Palatini potential}). Likewise, as one can confirm from Eqs.~\eqref{cutoff metric} and \eqref{cutoff Palatini}, the cutoff scales in the two scenarios at a plateau have the same dependence on the coupling $\xi$ as $\Lambda\sim \Mpl/\sqrt{\xi}$. Therefore, regardless of the choice of gravitational formulations, one finds that the Hubble scale is much smaller than the cutoff
\bae{
	\eval{\frac{H}{\Lambda}}_{\xi\phi_{\uJ i}^2/\Mpl^2\gg 1}\sim\sqrt{\frac{\lambda}{\xi}}\ll 1,
}
indicating that there is no unitarity violation during inflation.

Next, we consider the unitarity issue during reheating. Due to the rich interactions between the Higgs and the Standard Model particles, the potential energy of the inflaton (Higgs) is expected to be efficiently transferred to radiation as \emph{preheating}. Thus, the excited particles can have the momentum of $k\sim U^{1/4}_\uend$ rather than the Hubble scale, which has been indeed confirmed in Refs.~\cite{Ema:2016dny,Ema:2021xhq}. Here $U_\mathrm{end}$ denotes the inflatons' potential energy at the end of inflation. Accordingly, the analytical criterion for the unitarity violation is given by
\bae{
	U_\uend\gtrsim\eval{\Lambda^4}_{\phi_{\uJ i}^2\to0}.
}
It says that the perturbative unitarity is violated when the potential height $U_\uend$ exceeds the cutoff scale evaluated near the origin. In the case of the Higgs inflation, the potential is approximated as $U\sim \frac{\lambda\Mpl^4}{\xi^2}$ and thus the criterion of the unitarity violation reads
\bae{\label{metric unitarity}
	\mathrm{metric}&:\quad \eval{\frac{U_\uend}{\Lambda^4}}_{\phi_{\uJ i}^2\to0}\sim 10^3\lambda\xi^2,\\
	\label{Palatini unitarity}\mathrm{Palatini}&:\quad \eval{\frac{U_\uend}{\Lambda^4}}_{\phi_{\uJ i}^2\to0}\sim 10^2\lambda.
}
It should be noted that the potential used above is a value at a plateau, so the actual potential height at the end of inflation should be a bit lower and the criterion gets milder. However, the result would not change significantly.

The $\xi$ dependence in Eq.~\eqref{metric unitarity} clearly signals the unitarity violation during the preheating in the metric-Higgs inflation because the potential energy exceeds the cutoff in the large $\xi$ limit. This obviously requires UV-completion for accurate predictions of inflationary observables. As for the Palatini-Higgs inflation, the criterion~\eqref{Palatini unitarity} is free from $\xi$ and thus the unitarity violation may not be problematic practically, though it should be confirmed in a detailed investigation.

\subsubsection{UV-completion in the metric-Higgs inflation}\label{sec: Higgs-R2}

In the metric-Higgs inflation, several attempts of the UV-completion have been made so far~\cite{Lerner:2010mq,Giudice:2010ka,Barbon:2015fla,Salvio:2015kka,Ema:2017rqn,Gorbunov:2018llf,Lee:2018esk,Ema:2020zvg,Lee:2021dgi}. In particular, it is found in Ref.~\cite{Ema:2020zvg} that the Higgs inflation can be nicely understood as a nonlinear sigma model with a UV-completing sigma-field identified as a scalaron in $R^2$ operator which has to be added to renormalize one-loop divergences~\cite{tHooft:1974toh,Donoghue:1995cz}. The approach to the UV completion becomes clear once written as a nonlinear sigma model, and it is easy to confirm that the local conformal symmetry is respected in a resulting UV theory as we review Ref.~\cite{Ema:2020zvg} in the following.

We start with the action in the Jordan frame:
\bae{
    S=\int\dd[4]{x} \sqrt{-g_\uJ}\left[\frac{\Mpl^2}{2}\left(1+\xi\frac{\phi_{\uJ i}^2}{\Mpl^2}\right)R_\uJ-\frac{1}{2}g_\uJ^{\mu\nu}\partial_\mu\phi_{\uJ i} \partial_\nu\phi_{\uJ i}-\frac{\lambda}{4}(\phi_{\uJ i}^2)^2\right],
}
where we take the covariant gauge for the Higgs fields. Let us now introduce an unphysical scalar field $\Phi_\uJ$ called the conformal mode as a scaling factor of the metric through
\bae{
    g_{\uJ\mu\nu}\to
    g_{\mu\nu}=\frac{6\Mpl^2}{\Phi_\uJ^2}g_{\uJ\mu\nu}.
}
By redefining the Higgs fields as 
\bae{
    \phi_{\uJ i}\to\phi_i=\frac{\Phi_\uJ}{\sqrt{6}\Mpl}\phi_{\uJ i},
}
the action of the Higgs inflation can be rewritten in terms of $\Phi$ and $\phi_i$ as
\bae{\label{Eq: NLSM}
    S=\int\dd[4]{x} \sqrt{-g}\left[\frac{1}{12}(\Phi_\uJ^2+6\xi\phi_i^2)R-\frac{1}{2}g^{\mu\nu}\partial_\mu\phi_i \partial_\nu\phi_i+(6\xi+1)\frac{\phi_i}{\Phi_\uJ}g^{\mu\nu}\partial_\mu\phi_i \partial_\nu \Phi_\uJ \right.\nonumber\\
    \left. +\frac{1}{2}\left\{1-(6\xi+1)\frac{\phi_i^2}{\Phi_\uJ^2}\right\}g^{\mu\nu}\partial_\mu \Phi_\uJ \partial_\nu \Phi_\uJ-\frac{\lambda}{4}(\phi_i^2)^2 \right],
}
which can be viewed as a nonlinear sigma model with a curved field-space spanned by $\phi_i$ and $\Phi_\uJ$. In fact it can be rewritten in an ``apparently-flat'' way by introducing the $\sigma$ field defined by
\bae{
    \sigma=\frac{1}{2}\bqty{\sqrt{\Phi^2-2(6\xi+1)\phi_i^2}-\Phi},
}
with the redefinition of $\Phi_\uJ$
\bae{
    \Phi_\uJ=\frac{1}{2}\bqty{\sqrt{\Phi^2-2(6\xi+1)\phi_i^2}+\Phi},
}
as
\bae{\label{Eq: NLSM with h}
    S=\int\dd[4]{x} \sqrt{-g}\left[\frac{1}{12}(\Phi^2-\phi_i^2-\sigma^2)R+\frac{1}{2}(\partial_\mu \Phi)^2-\frac{1}{2}(\partial_\mu\phi_i)^2-\frac{1}{2}(\partial_\mu \sigma)^2-\frac{\lambda}{4}(\phi_i^2)^2\right],
}
where $\sigma=\sigma(\phi_i,\Phi)$ is a function of $\phi_i$ and $\Phi$. 
Note that the theory contains four real degrees of freedom corresponding to the number of the Higgs fields, which is understood as the following five-dimensional hypersurface in $\mathbb{R}^{(1,5)}$ with one conformal gauge degree of freedom $\Phi$:
\bae{\label{Eq: constraint hypersurface}
    \frac{6\xi+1}{2}\phi_i^2+\left(\sigma+\frac{\Phi}{2}\right)^2=\frac{\Phi^2}{4}, \quad \text{in $(\Phi,\phi_i,\sigma)\in \mathbb{R}^{(1,5)}$}.
}
Once the Higgs inflation is written in the form of Eq.~\eqref{Eq: NLSM with h}, it is clear that the UV-completion can be realized by promoting $\sigma$ to a dynamical field. The constraint~\eqref{Eq: constraint hypersurface} can be instead introduced as a potential constraint as
\bae{\label{Eq: LSM}
    S_\alpha=\int\dd[4]{x} \sqrt{-g}\Biggl[\frac{1}{12}(\Phi^2-\phi_i^2-\sigma^2)R+\frac{1}{2}(\partial_\mu \Phi)^2 -\frac{1}{2}(\partial_\mu \phi_i)^2  -\frac{1}{2}(\partial_\mu \sigma)^2 \nonumber\\
    -\frac{\lambda}{4}(\phi_i^2)^2 -\frac{1}{144\alpha}\left[\frac{\Phi^2}{4}-\left(\sigma+\frac{\Phi}{2}\right)^2-\frac{6\xi+1}{2}\phi_i^2\right]^2 \Biggr],
}
which reduces to the original Higgs inflation~\eqref{Eq: NLSM with h} in the limit of small $\alpha$. This action is actually equivalent in the scalar part to the original one with the $\alpha R^2$ term in the Jordan frame as this term introduces the dynamical scalaron $\sigma$.\footnote{See also, e.g., Refs.~\cite{Kannike:2015apa,Salvio:2016vxi,He:2018gyf,He:2018mgb,He:2020ivk,He:2020qcb} for the Higgs-$R^2$ theory in the metric approach and Refs.~\cite{Gialamas:2019nly,Gialamas:2020vto,Gialamas:2021enw} in the Palatini formulation.} Note that the size of $\alpha$ is generically of order $\calO(\xi^2)$ once we assume the large non-minimal coupling $\xi$ ~\cite{Ema:2020zvg}.

Here it should be stressed that the action~\eqref{Eq: LSM} has the local conformal symmetry as it is invariant under the following transformation:
\bae{
 g_{\mu\nu}\to\tilde{g}_{\mu\nu}=\ee^{-2\rho(x)}g_{\mu\nu}, \qquad \Phi\to\tilde{\Phi}=\ee^{\rho(x)}\Phi, \qquad \phi_i\to\tilde{\phi}_i=\ee^{\rho(x)}\phi_i, \qquad \sigma\to\tilde{\sigma}=\ee^{\rho(x)}\sigma,
}
with a scalar function $\rho(x)$. With use of the gauge symmetry, the ghost-like field $\Phi$ can be killed by taking a specific gauge choice, so that it does not cause any problem. In discussing the unitarity problem in the UV-complete theory, it is useful to adopt the Einstein frame because the gravity sector is merely given by the Einstein--Hilbert form and the perturbation theory is known to be valid below the Planck scale. The Einstein frame expression can be obtained by fixing the gauge through
\bae{\label{Eq. gauge condition metric-Higgs}
\Phi^2-\phi_i^2-\sigma^2=6\Mpl^2.
}
Since there is no $\xi$-dependence in the above gauge condition, both the kinetic and the potential sector do not generate a lower cutoff than the Planck scale.\footnote{Note that the kinetic and potential terms originally consist only of the renormalizable four-dimensional operators unless the gauge condition~\eqref{Eq. gauge condition metric-Higgs}.} In this sense, the action~\eqref{Eq: LSM} is free from the unitarity problem.\footnote{Practically, it is not necessary to be valid above the Planck scale where the spin-2 sector becomes significant.}

The direct generalization of this approach to the Palatini-Higgs inflation, however, would not work because the $R^2$ term does not give rise to a new physical scalar degree of freedom (DoF) in the Palatini gravity. Indeed any theory whose gravity sector is given by a function $f(R,S)$ of the Ricci scalar $R$ and some scalar(s) $S$ can be simplified to the Einstein gravity without deriving any additional scalar DoF as
\bae{
	S&\supset\frac{1}{2}\int\dd[4]{x}\sqrt{-g_\uJ}f(R_\uJ,S)=\frac{1}{2}\int\dd[4]{x}\sqrt{-g_\uJ}\bqty{f(\omega,S)+(R_\uJ-\omega)\partial_\omega f(\omega,S)} \nonumber \\
	&=\int\dd[4]{x}\sqrt{-g}\bqty{\frac{1}{2}\Mpl^2R-3\kappa\Mpl^2\frac{\Box\sqrt{\partial_\omega f(\omega,S)}}{\sqrt{\partial_\omega f(\omega,S)}}-\Mpl^4\frac{\omega\partial_\omega f(\omega,S)-f(\omega,S)}{2(\partial_\omega f(\omega,S))^2}},
}
with the rescaling $g_{\uJ\mu\nu}=\frac{\Mpl^2}{\partial_\omega f(\omega,S)}g_{\mu\nu}$ and an auxiliary field $\omega$~\cite{DeFelice:2010aj}. Note that $\omega$ is non-dynamical because $\kappa=0$ in the Palatini case~\cite{Edery:2019txq},\footnote{Though $\omega$ itself is non-dynamical in the Palatini case, the Euler--Lagrange constraint on $\omega$ gives rise to higher-order derivatives in the Higgs field (see, e.g., Refs.~\cite{Antoniadis:2018ywb,Tenkanen:2019jiq,Lykkas:2021vax}) and they may be understood as an emergence of a new DoF.} while the metric formalism $\kappa=1$ makes $\omega$ dynamical and causes the scalaron $\sigma$ after some field redefinition.
In the next section, we investigate the UV completion in the Palatini-Higgs model by an additional scalar, not restricting ourselves to $R^2$, but in fact, we find the simple flattening of the field-space by a new scalar does not UV-complete the theory.

\section{UV-(in)completion}\label{sec3}

We have reviewed phenomenological aspects of the metric-Higgs and Palatini-Higgs inflation and seen that the metric-Higgs inflation should be UV-completed for precise predictions while the Palatini-Higgs one may be by itself predictive because energy scales are smaller than the cutoff. From a theoretical point of view, however, a new physics must intervene even in the Palatini-Higgs scenario by the Planck scale. Therefore it is interesting to investigate what happens above the cutoff scale in the two scenarios.

The essence of the UV completion in the metric-Higgs-$R^2$ scenario reviewed in Sec.~\ref{sec: Higgs-R2} is to interpret the curved field-space as a hypersurface embedded into the one-higher-dimensional space with the scalaron introduced by the $R^2$ term.
We first show that the direct embedding with a new scalar, which does not necessarily originate from the $R^2$ term, can indeed UV complete the metric-Higgs model.
This approach seems helpful also for the Palatini-Higgs scenario, in which the $R^2$ term does not yield a new scalar DoF.
However, in the following section, we see that the direct embedding does not work in the Palatini scenario.
In Sec.~\ref{sec3-2}, we interpret these facts in the context of the local conformal symmetry.

\subsection{Embedding $\calM_4$ into $\mathbb{R}^5$}\label{sec3-1}

The low cutoff originates in the non-zero curvature of the curved field-space $\calM_4$. Therefore we consider introducing an additional scalar field which flattens the field-space. We will first provide an easy-to-use geometrical way to embed the four-dimensional field-space $\calM_4$ into a five-dimensional flat field-space $\mathbb{R}^5$, which is applicable to both the metric-Higgs and Palatini-Higgs inflation.\footnote{It should be noted that the idea of flattening the field-space was first introduced in Ref.~\cite{Giudice:2010ka} in which an added scalar field can be interpreted as a $\sigma$-meson in the context of the nonlinear sigma model and has been further developed in a rigorous manner in Ref.~\cite{Ema:2020zvg}.} We will then demonstrate that this approach hardly UV-completes the Palatini-Higgs inflation while a new scalar field uplifts the cutoff to the Planck scale in the metric-Higgs inflation.

We first note that $\calM_4$ should be spherically symmetric because of the symmetry of the Higgs. In fact, one finds that the line element of $\calM_4$ in the Cartesian coordinate expression~\eqref{field-space},
\bae{
    \dd{s^2(\calM_4)}=G_{ij}(\phi_\uJ)\dd{\phi_{\uJ i}}\dd{\phi_{\uJ j}} \qc
    G_{ij}(\phi_\uJ)=\frac{1}{1+\xi\frac{\phi_{\uJ k}^2}{\Mpl^2}}\pqty{\delta_{ij}+\frac{6\kappa\xi^2\frac{\phi_{\uJ i}\phi_{\uJ j}}{\Mpl^2}}{1+\xi\frac{\phi_{\uJ k}^2}{\Mpl^2}}},
}
can be rewritten in the spherical coordinate as
\bae{\label{eq: ds in Riemannian}
	\dd{s^2(\calM_4)}=\frac{1+\xi(6\kappa\xi+1)\frac{r_\uJ^2}{\Mpl^2}}{\qty(1+\xi\frac{r_\uJ^2}{\Mpl^2})^2}\dd{r_\uJ^2}+\frac{r_\uJ^2}{1+\xi\frac{r_\uJ^2}{\Mpl^2}}\dd{\Omega_3}\eqqcolon F^2(r_\uJ)\dd{r_\uJ^2}+r^2(r_\uJ)\dd{\Omega_3}
}
where $r_\uJ^2=\phi_{\uJ i}^2$ and $r^2\dd{\Omega_3}$ is the angular line element of the three-dimensional $r$-sphere. $\kappa=1$ for the metric formalism and $\kappa=0$ for the Palatini.
This line element can be deformed as
\bae{
	\dd{s^2(\calM_4)}=\pqty{F^2(r_\uJ)-\pqty{
	r^\prime(r_\uJ)}^2}\dd{r_\uJ^2}+\pqty{\dd{
	r(r_\uJ)}}^2+r^2(r_\uJ)\dd{\Omega_3}.
}
Therefore, if $F^2(r_\uJ)-\pqty{r^\prime(r_\uJ)}^2$ is always non-negative which holds true both in the metric-Higgs and the Palatini-Higgs models,
it can be understood as a hypersurface embedded in the flat five-dimensional field-space $\mathbb{R}^5$ with the line element
\bae{
    \dd{s^2(\mathbb{R}^5)}=\dd{z^2}+\dd{r^2}+ r^2\dd{\Omega_3},
}
as
\bae{\label{hypersurface}
    \dd{s^2(\calM_4)}=\eval{\dd{s^2(\mathbb{R}^5)}}_{z=z_\sur(
    r^2)} \qc
    z_\sur(
    r^2)=\int_0^{r_\uJ(
    r^2)}\sqrt{F^2(r_\uJ^\prime)-\left(\dv{
    r(r_\uJ^\prime)}{r_\uJ^\prime}\right)^2}\dd{r_\uJ^\prime}.
}
Note that $r_\uJ$ and $r$ are related by
\bae{
	r^2=\frac{r_\uJ^2}{1+\xi\frac{r_\uJ^2}{\Mpl^2}} \qc 
	\Leftrightarrow \quad r_\uJ^2=\frac{
	r^2}{1-\xi\frac{
	r^2}{\Mpl^2}}.
}

From this viewpoint, the original (four-scalar) Higgs inflation with a non-canonical kinetic term is equivalent to a canonical five scalars' theory constrained on the hypersurface~\eqref{hypersurface}. 
The unitarity violation in the Higgs inflation can then originate from a ``unphysical'' constraint onto this hypersurface, and the five-scalar theory with a ``physical'' hypersurface constraint can be a possible candidate of the UV theory of the Higgs inflation.
For example, let us introduce the hypersurface constraint by the additional $r$'s quartic coupling,
\bae{\label{eq: additional potential}
    V_\mathrm{sur}(
    r^2,z)=\frac{\tilde{\lambda}}{4}\pqty{r^2-r^2_\sur(z)}^2,
}
where $r^2_\sur(z)$ is the inverse of $z_\sur(r^2)$~\eqref{hypersurface}.
This potential restricts $r$ onto the hypersurface with the effective mass squared,
\bae{
	\eval{\dv[2]{V_\sur}{r}}_{r^2=r^2_\sur(z)}=2\tilde{\lambda}
	r_\sur^2(z).
}
Therefore, while the Higgs inflation is reproduced in the lower energy region than $\sim\sqrt{2\tilde{\lambda}r^2_\sur(z)}$, this potential reduces to a mere renormalizable coupling in the UV regime as long as the hypersurface term $r^2_\sur(z)$ does not contain any violent higher-dimensional operators.

In the following subsections, as a possible UV-completion of the Higgs inflation, we investigate the five canonical scalars' theory given by 
\bae{\label{eq: S5}
	S_5=\int\dd[4]{x}\sqrt{-g}\left[\frac{\Mpl^2}{2}R-\frac{1}{2}\pqty{(\partial z)^2+(\partial\phi_i)^2}-\frac{\lambda}{4}(\phi_i^2)^2 %+
	-V_\sur(\phi_i^2,z)\right],
}
where we adopt the Cartesian coordinate $\phi_i$
which satisfies $r^2=\phi_i^2$ in the five-dimensional field-space $\mathbb{R}^5$.
Here we used the relation
\bae{
    	\frac{\lambda}{4}\frac{(\phi_{\uJ i}^2)^2}{\pqty{1+\xi\frac{\phi_{\uJ i}^2}{\Mpl^2}}^2}=\frac{\lambda}{4}(\phi_i^2)^2
}
to rewrite the original Higgs quartic potential in terms of $\phi_i$.\footnote{The gauge sector also consists only of dimension-four operators in terms of $\phi_i$ as discussed in Sec.~\ref{sec: covariant gauge}.} 
As the field-space is now flat in five dimensions, it is expected that the perturbative unitarity can be restored up to the Planck scale (or the Landau pole for quartic coupling constants $\lambda$ or $\tilde{\lambda}$) unless the hypersurface term $r_\sur^2(z)$ gives rise to a lower cutoff scale.

\subsubsection{Metric-Higgs inflation}\label{sec3-1-1}

We first analyse the five scalars' theory in the metric-Higgs inflation. The hypersurface can be obtained by integrating Eq.~\eqref{hypersurface} as
\bae{\label{eq: zrho in Riemannian}
    z_\sur(
    r^2)=\frac{\Mpl}{\sqrt{\xi}}\Biggl[\sqrt{6\xi+1}\pqty{\sinh^{-1}\sqrt{\frac{6\xi+1}{1-\xi\frac{
    r^2}{\Mpl^2}}}-\sinh^{-1}\sqrt{6\xi+1}} \nonumber \\
    \quad-\sqrt{(6\xi+1)+\pqty{1-\xi\frac{
    r^2}{\Mpl^2}}}+\sqrt{(6\xi+1)+1}\Biggr].
}
In the large $\xi$ limit, the hypersurface asymptotes to the following form
\bae{\label{eq: asymptotic hypersurface in Riemannian}
    \sqrt{\frac{\xi}{6\xi+1}}\frac{z_\sur(
    r^2)}{\Mpl}\sim-\frac{1}{2}\log\pqty{1-\xi\frac{
    r^2}{\Mpl^2}} \qc 
    \Leftrightarrow \quad \xi\frac{
    r^2_\sur(z)}{\Mpl^2}\sim1-\exp\pqty{-2\sqrt{\frac{\xi}{6\xi+1}}\frac{z}{\Mpl}},
}
and the potential reads
\bae{\label{eq: V in metric}
    V(\phi_i^2,z)\sim\frac{\lambda}{4}(\phi_i^2)^2+\frac{\tilde{\lambda}}{4}\bqty{\phi_i^2-\frac{\Mpl^2}{\xi}\pqty{1-\exp\pqty{-2\sqrt{\frac{\xi}{6\xi+1}}\frac{z}{\Mpl}}}}^2.
}
In the above potential, a plateau which is suitable for slow-roll is realized in the $z$-direction where $r$ can be approximated as a constant. So the new particle $z$ becomes nearly massless and it plays a role of the inflaton. On the other hand, along the $z$-direction, $r^2_\sur(z)$ is well approximated by $\Mpl^2/\xi$ without depending on $z$ and hence the radial direction of the Higgs fields $(r^2=\phi^2_i)$ obtains the mass from the potential
\bae{
    V(\phi_i^2,z)\sim\frac{\lambda}{4}(\phi_i^2)^2+\frac{\tilde{\lambda}}{4}\pqty{\phi_i^2-\frac{\Mpl^2}{\xi}}^2.
}
By differentiating twice with respect to the radial component and substituting the approximated value $r^2_{\mathrm{sur}}=\Mpl^2/\xi$, the mass can be read off as
\bae{
\eval{m_r^2}_{r=r_\sur}= \eval{\partial_r^2 V}_{r=r_\sur}\sim \Bqty{3(\lambda+\tilde{\lambda})r^2-\tilde{\lambda}\frac{\Mpl^2}{\xi}}_{r=r_\sur}=\frac{3\lambda+2\tilde{\lambda}}{\xi}\Mpl^2,
}
corresponding to the cutoff scale~\eqref{eq: cutoff during inflation} without the $z$ particle. As it is larger enough than the Hubble scale, $H^2\sim\eval{\frac{V}{3\Mpl^2}}_{r_\sur}\sim\frac{\lambda}{12\xi^2}\Mpl^2$, this inflation theory can be treated as an effective single-field model. With the hypersurface constraint
\bae{
    r^2=r_{\mathrm{sur}}^2(z)\sim\frac{\Mpl^2}{\xi}\left[1-\exp\left(-2\sqrt{\frac{\xi}{6\xi+1}}\frac{z}{\Mpl}\right)\right],
}
the effective potential in terms of $z$ along the hypersurface is given by
\bae{
    \eval{V(\phi_i^2,z)}_{r_\sur}=\frac{\lambda}{4}r_\sur^4(z)\sim\frac{\lambda\Mpl^4}{4\xi^2}\bqty{1-\exp\pqty{-2\sqrt{\frac{\xi}{6\xi+1}}\frac{z}{\Mpl}}}^2,
}
which in fact asymptotes to the original potential~\eqref{metric potential} in the inflationary regime $z\gtrsim\Mpl$.
Therefore the five-scalar model~\eqref{eq: S5} can explain the origin of the low cutoff scale $\Lambda\sim\Mpl/\sqrt{\xi}$ without spoiling the inflationary phenomenology.

The asymptotic form of the hypersurface $r^2_\sur$~\eqref{eq: asymptotic hypersurface in Riemannian} also shows that the Taylor expansion of the potential around the origin does not yield any violent higher-dimensional operators which are proportional to $\xi^n$ with positive power $n$:
\bae{
    V_{\mathrm{sur}}(\phi_i^2,z)\simeq\frac{\tilde{\lambda}}{4}\left(\frac{2\Mpl^2}{3\xi^2}z^2+(\phi_i^2)^2-\frac{4\Mpl}{3\sqrt{6}\xi^2}z^3-\frac{4\Mpl}{\sqrt{6}\xi}\phi_i^2 z+\frac{2}{3\xi}\phi_i^2z^2+\frac{7}{27\xi^2}z^4 \right. \nonumber \\
    \left. -\frac{4}{9\sqrt{6}\xi\Mpl}\phi_i^2z^3-\frac{2}{9\sqrt{6}\xi\Mpl}z^5+\frac{1}{27\xi\Mpl^2}\phi_i^2 z^4+\cdots\right).\label{Eq. Vsur around metric}
}
We remark that this expanded potential is confirmed to be equivalent to the leading order expression of the full hypersurface constraint~\eqref{eq: zrho in Riemannian}. From the potential~\eqref{Eq. Vsur around metric}, it turns out that the theory is healthy as well around the origin up to the Planck scale. The low cutoff scale $\Lambda\sim\Mpl/\xi$ in the original theory~\eqref{eq: cutoff around origin in metric} can be now understood as the mass of the new particle $z$: one finds that the Hessian of the potential~\eqref{eq: V in metric} around the origin is zero except for
\bae{
\eval{\partial_z^2V}_{\phi_i=0,z=0}=\frac{\tilde{\lambda}}{3\xi^2}\Mpl^2.
}
The new scalar $z$ has mass $\sim\sqrt{\tilde{\lambda}}\Mpl/\xi$ and should be excited above this scale.

\subsubsection{Palatini-Higgs inflation}\label{sec3-1-2}

Let us turn to the Palatini-Higgs inflation. The hypersurface constraint can be obtained from Eq.~\eqref{hypersurface} as
\bae{\label{eq: zrho in Palatini}
    \frac{\sqrt{\xi}}{\Mpl}z_\sur(r^2)=\sqrt{2}-\sqrt{2-\frac{\xi r^2}{\Mpl^2}}-\sinh^{-1}[1]+\sinh^{-1}\sqrt{\frac{1}{1-\frac{\xi r^2}{\Mpl^2}}}.
}
Although the hypersurface cannot be further simplified in the large $\xi$ limit as opposed to the metric case, one can verify that the potential is flat in the large $z$ region and the new particle $z$ can act as the inflaton. 
In the large $z$ region, the hypersurface asymptotes to
\bae{
    \frac{1}{1-\frac{\xi}{\Mpl^2}r^2_{\mathrm{sur}}}\sim\sinh^2{\frac{\sqrt{\xi}}{\Mpl}z_{\mathrm{sur}}}\qc \Rightarrow\quad r^2_{\mathrm{sur}}\sim \frac{\Mpl^2}{\xi}\pqty{1-4\ee^{-2\sqrt{\xi}z/\Mpl}},
}
so that the effective potential along the hypersurface behaves as
\bae{
    \eval{V(\phi_i^2,z)}_{r_\sur}=\frac{\lambda}{4}r_\sur^4(z)\sim\frac{\lambda\Mpl^4}{4\xi^2}\pqty{1-8\ee^{-2\sqrt{\xi}z/\Mpl}},
}
which asymptotes to the original potential~\eqref{Palatini potential}.
Meanwhile the approximately constant constraint $r^2_\sur\sim \Mpl^2/\xi$ gives rise to the mass of Higgs fields as $\eval{m_r^2}_{r_\sur}\sim\frac{3\lambda+2\tilde{\lambda}}{\xi}\Mpl^2$ which again explains the cutoff during inflation~\eqref{eq: cutoff during inflation}.

However, the cutoff around the origin is not actually explained simply by the $z$ particle in contrast to the metric case. The expansion of the potential \eqref{eq: additional potential} around the origin reads
\bae{
    V_{\mathrm{sur}}(\phi_i^2,z)=\frac{\tilde{\lambda}}{4}\left(\frac{2\Mpl^2}{\xi}z^2-\frac{3\Mpl}{\sqrt{2\xi}}z^3-\frac{2\sqrt{2}\Mpl}{\sqrt{\xi}}\phi_i^2 z+(\phi_i^2)^2
    %+\frac{3}{2}\tilde{\bm{\phi}}^2z^2
    +\frac{3}{2}\phi_i^2z^2
    +\frac{43}{48}z^4 \right. \nonumber \\
    \left. -\frac{1}{3\sqrt{2}}\frac{1}{\Mpl/\sqrt{\xi}}\phi_i^2z^3-\frac{1}{8\sqrt{2}}\frac{1}{\Mpl/\sqrt{\xi}}z^5
    +\cdots\right). 
}
The higher-dimensional operators in the second line are suppressed only by the lower cutoff $\Mpl/\sqrt{\xi}$.
This can be understood that the five scalars' theory is not valid up to the Planck scale even though the field-space is flattened by the new particle $z$ with its mass
\bae{
\eval{\partial_z^2V}_{\phi_i=0,z=0}=\frac{\tilde{\lambda}}{\xi}\Mpl^2.
}
In other words, it turns out that the simple embedding to the flat field-space does not UV-complete the Palatini-Higgs inflation.

\subsection{Conformal symmetry}\label{sec3-2}

In the previous subsection, we found that one additional scalar field introduced to flatten the field-space hardly UV-completes in the Palatini-Higgs inflation while the same approach leads to a successful scenario in the metric-Higgs inflation. 
In this subsection, we interpret the origin of the (unavoidable) field-space curvature of the Palatini-Higgs model in terms of the local conformal symmetry, comparing it with the metric one reviewed in Sec.~\ref{sec: Higgs-R2}.

In the Palatini formalism, the local conformal transformation is defined by the change of the metric while the connection is left unaffected:
\bae{
    g_{\mu\nu}\to\tilde{g}_{\mu\nu}=\ee^{-2\rho(x)}g_{\mu\nu}, \qquad S\to\tilde{S}=\ee^{\rho(x)}S, \qquad \Gamma^\tau_{\mu\nu}\to \tilde{\Gamma}^\tau_{\mu\nu}=\Gamma^\tau_{\mu\nu},
}
with arbitrary scalar(s) $S$.
Since the Ricci tensor is a function only of the connection, the Ricci scalar $R=g^{\mu\nu}R_{\mu\nu}(\Gamma)$ transforms covariantly under the local conformal transformation, $R\to\tilde{R}=\ee^{2\rho(x)}R$. Therefore the non-minimal coupling $\sqrt{-g}S^2R$ is conformally invariant by itself. Meanwhile, the kinetic term of a scalar does not exhibit the conformal invariance by itself and thus one is required to introduce the covariant derivative. In the Palatini formalism, the metric compatibility $\nabla_\mu g_{\alpha\beta}=0$ is discarded and a geometrical vector field $Q_\mu$ called the non-metricity~\cite{Kleyn:2004yj} is naturally introduced:
\bae{
    Q_\mu\coloneqq- g^{\alpha\beta}\nabla_\mu g_{\alpha\beta}.
}
This non-metricity can play the role of the conformal gauge field implementing the covariant derivative $D_\mu\coloneqq \partial_\mu -\frac{1}{8}Q_\mu$ for scalar fields as one can check its transformation law $Q_\mu\to\tilde{Q}_\mu=Q_\mu+8\partial_\mu\rho$~\cite{Iosifidis:2018zwo}.

We then note that the usual kintetic term of Higgs fields can be understood as the (conformal) covariant one in the large mass limit of the non-metricity:
\bae{
    -\frac{1}{2}g^{\mu\nu}\partial_\mu \phi_{\uJ i}\partial_\nu\phi_{\uJ i}=-\frac{1}{2}g^{\mu\nu}D_\mu\phi_{\uJ i} D_\nu\phi_{\uJ i} -\eval{\frac{1}{2}\Lambda^2 Q_{\mu} Q^\mu}_{\Lambda\to \infty}.
}
The mass term of the non-metricity can be realized by the gauge fixing of the conformally invariant action
\bae{\label{eq: conformal action}
    S_{\conf}=\int\dd[4]{x} \sqrt{-g}\left[
    \frac{1}{2}\pqty{c
    \Phi_\uJ^2+\xi
    \phi_{\uJ i}^2}R
    -\frac{1}{2}g^{\mu\nu}D_\mu
    \Phi_\uJ D_\nu 
    \Phi_\uJ-\frac{1}{2}g^{\mu\nu}D_\mu \phi_{\uJ i} D_\nu \phi_{\uJ i} 
    -\frac{\lambda}{4}(
    \phi_{\uJ i}^2)^2\right],
}
where $\phi_{\uJ i}$ denotes the Higgs fields, $\Phi_\uJ$ represents an additional scalar field which will be killed by the gauge fixing, and $c$ is its non-minimal coupling constant.
Fixing the local conformal symmetry by the gauge condition $c \Phi_\uJ^2=\Mpl^2$, one finds
\bae{
    S_{\conf}=\int\dd[4]{x} \sqrt{-g}\left[\frac{\Mpl^2}{2}\left(1+\xi\frac{
    \phi_{\uJ i}^2}{\Mpl^2}\right)R 
   -\frac{1}{2}\frac{\Mpl^2}{64c}Q_\mu Q^\mu-\frac{1}{2}g^{\mu\nu}D_\mu \phi_{\uJ i} D_\nu \phi_{\uJ i} 
    -\frac{\lambda}{4}(
    \phi_{\uJ i}^2)^2\right],
}
which reduces to the original Higgs-inflation action in the large $Q_\mu$-mass limit, $c\to0$.

Contrary to the metric-Higgs model, the kinetic term of the conformal action~\eqref{eq: conformal action} is already diagonalized without introducing the scalaron.
Accordingly, the field-space curvature of the Palatini-Higgs model can be understood to be caused by the gauge fixing as follows.
Once the local conformal symmetry is explicitly exhibited, being similar to the metric-Higgs case~\eqref{Eq. gauge condition metric-Higgs}, the Einstein frame is easily taken by the gauge fixing, 
\bae{\label{eq: Einstein gauge}
    c
    \Phi_\uJ^2+\xi
    \phi_{\uJ i}^2=\Mpl^2.
}
This gauge is specifically realized, e.g., by the reparametrization
\bae{
    \bce{
        \dps
        \Phi_\uJ=\frac{\Mpl}{\sqrt{c}}\cos\frac{\varphi_1}{\Mpl}, \\[10pt]
        \dps
        \phi_{\uJ 1}=\frac{\Mpl}{\sqrt{\xi}}\sin\frac{\varphi_1}{\Mpl}\cos\frac{\varphi_2}{\Mpl}, \\[10pt]
        \dps
        \phi_{\uJ 2}=\frac{\Mpl}{\sqrt{\xi}}\sin\frac{\varphi_1}{\Mpl}\sin\frac{\varphi_2}{\Mpl}\cos\frac{\varphi_3}{\Mpl}, \\[10pt]
        \dps
        \phi_{\uJ 3}=\frac{\Mpl}{\sqrt{\xi}}\sin\frac{\varphi_1}{\Mpl}\sin\frac{\varphi_2}{\Mpl}\sin\frac{\varphi_3}{\Mpl}\cos\frac{\varphi_4}{\Mpl}, \\[10pt]
        \dps
        \phi_{\uJ 4}=\frac{\Mpl}{\sqrt{\xi}}\sin\frac{\varphi_1}{\Mpl}\sin\frac{\varphi_2}{\Mpl}\sin\frac{\varphi_3}{\Mpl}\sin\frac{\varphi_4}{\Mpl}.
    }
}
The kinetic terms then read, in the small $c$ limit,
\bae{
    &-\frac{1}{2}(D_\mu 
    \Phi_\uJ)^2-\frac{1}{2}(D_\mu\phi_{\uJ i})^2\to-\frac{1}{2c}\pqty{(\partial_\mu\varphi_1)\sin\frac{\varphi_1}{\Mpl}+\frac{\Mpl}{8}Q_\mu\cos\frac{\varphi_1}{\Mpl}}^2 \nonumber \\
    &\quad-\frac{1}{2\xi}
    \left[(\partial_\mu\varphi_1)^2\sec^2\frac{\varphi_1}{\Mpl}+(\partial_\mu\varphi_2)^2\sin^2\frac{\varphi_1}{\Mpl}+(\partial_\mu\varphi_3)^2\sin^2\frac{\varphi_1}{\Mpl}\sin^2\frac{\varphi_2}{\Mpl} \right. \nonumber \\
    &\qquad\left.+(\partial_\mu\varphi_4)^2\sin^2\frac{\varphi_1}{\Mpl}\sin^2\frac{\varphi_2}{\Mpl}\sin^2\frac{\varphi_3}{\Mpl}\right].
}
The first term vanishes due to the Euler--Lagrange constraint on $Q_\mu$, and the rest terms give rise to the field-space curvature $\calR_G=\frac{3(5+3\cos\frac{2\varphi_1}{\Mpl})}{2}\frac{\xi}{\Mpl^2}$ for $\varphi_{1,2,3,4}$.
This curvature originates from the prolate gauge condition~\eqref{eq: Einstein gauge} with the short semi-minor axis $\sim\Mpl/\sqrt{\xi}$, and this is the origin of the low cutoff scale $\Lambda\sim\Mpl/\sqrt{\xi}$ in the Palatini-Higgs inflation. As the gauge condition is required for the Einstein frame picture, this low cutoff is unavoidable.

Let us remark that the successful metric-Higgs-$R^2$ model can be reproduced also in the Palatini formulation by simply generalizing its conformal picture~\eqref{Eq: LSM} to the Palatini approach as
\bae{
    S=\int\dd[4]{x}\sqrt{-g}\Biggl[\frac{1}{12}(
    \Phi^2-
    \phi_i^2-
    \sigma^2)R+\frac{1}{2}(D_\mu 
    \Phi)^2-\frac{1}{2}(D_\mu
    \phi_i)^2-\frac{1}{2}(D_\mu 
    \sigma)^2 \nonumber \\
    -\frac{\lambda}{4}(
    \phi_i^2)^2-\frac{1}{144\alpha}\bqty{\frac{
    \Phi^2}{4}-\pqty{
    \sigma+\frac{
    \Phi}{2}}^2-\frac{6\xi+1}{2}
    \phi_i^2}^2\Biggr].
}
The only difference $Q_\mu$ does not matter because it is integrated out as $Q_\mu\to0$ with a large enough mass $\sim\Mpl$ in the Einstein gauge fixing $\Phi^2-\phi_i^2-\sigma^2=6\Mpl^2$.

\section{Conclusions}\label{sec4}

In this paper, we have investigated the possibility of the UV-completion of the metric-Higgs and the Palatini-Higgs scenarios, both of which are not valid up to the Planck scale due to a curved field-space of four real-scalar Higgs with a smaller curvature radius than the Planck scale. In Sec.~\ref{sec3}, we have first presented an approach to embed the curved field-space into one-higher-dimensional flat space with a new scalar field, and then studied the five scalars' theory~\eqref{eq: S5} in both scenarios as a possible candidate of the UV-completion. We found that, in the metric formalism, the resulting five scalars' theory does not contain any violent higher-dimensional operators, so that the perturbation theory becomes valid during inflation and reheating. On the other hand, this direct embedding does not work in the Palatini-Higgs inflation and the cutoff scale remains unchanged as $\Mpl/\sqrt{\xi}$, even though the field-space is flattened by the new field.

These facts can be understood in the context of the local conformal symmetry. In the metric-Higgs scenario, the local conformal symmetry does not allow the kinetic term of Eq.~\eqref{Eq: NLSM} to be canonical but forces it to have the $\xi$-dependence. 
This situation rather enables us to eliminate the parameter $\xi$ in the non-minimal coupling and the kinetic term at all once via a suitable field inclusion. 
In the Palatini-Higgs inflation, however, the Ricci scalar differently responds to the local conformal symmetry than the metric formulation and the kinetic terms can be initially diagonalized because each term can hold the conformal invariance with the naturally-introduced conformal gauge field. Thus, the $\xi$-dependence of the non-minimal coupling and the kinetic term cannot be erased simultaneously and the gauge condition to move to the Einstein frame~\eqref{eq: Einstein gauge} unavoidably yields the low cutoff scale.

Our result shows that, while a single additional scalar field introduced to flatten the curved field-space successfully UV-completes in the metric-Higgs inflation, this embedding approach cannot uplift the low cutoff scales in the Palatini-Higgs case. Though the possibility of the UV-completion by multiple fields still remains, it seems doubtful because the infinite number of non-renormalizable operators should be regularized by the finite number of particles. Furthermore Ref.~\cite{Jinno:2019und} shows that the prediction of the Palatini-Higgs inflation is not steady against higher-dimensional operators. Thus the Palatini-Higgs model might be ``unnatural'' as a low-scale perturbative EFT. Another possibility is that the Palatini-Higgs inflation is UV-completed in a non-perturbative manner, which is worth investigating. 
We leave it as a future issue.

\acknowledgments

We are grateful to Keisuke Izumi, Kyohei Mukaida, and Shuichiro Yokoyama for helpful discussions.
Y.T. is supported by JSPS KAKENHI Grants 
No. JP19K14707 and No. JP21K13918.

\bibliographystyle{JHEP}
\bibliography{Bib}

\providecommand{\href}[2]{#2}\begingroup\raggedright\begin{thebibliography}{10}

\bibitem{Futamase:1987ua}
T.~Futamase and K.-i.~Maeda, \emph{{Chaotic Inflationary Scenario in Models
  Having Nonminimal Coupling With Curvature}},
  \href{https://doi.org/10.1103/PhysRevD.39.399}{\emph{Phys. Rev. D} {\bfseries
  39} (1989) 399}.

\bibitem{Bezrukov:2007ep}
F.L.~Bezrukov and M.~Shaposhnikov, \emph{{The Standard Model Higgs boson as the
  inflaton}}, \href{https://doi.org/10.1016/j.physletb.2007.11.072}{\emph{Phys.
  Lett. B} {\bfseries 659} (2008) 703}
  [\href{https://arxiv.org/abs/0710.3755}{{\ttfamily 0710.3755}}].

\bibitem{Barvinsky:2008ia}
A.O.~Barvinsky, A.Y.~Kamenshchik and A.A.~Starobinsky, \emph{{Inflation
  scenario via the Standard Model Higgs boson and LHC}},
  \href{https://doi.org/10.1088/1475-7516/2008/11/021}{\emph{JCAP} {\bfseries
  11} (2008) 021} [\href{https://arxiv.org/abs/0809.2104}{{\ttfamily
  0809.2104}}].

\bibitem{Einstein:1925:EFG}
A.~Einstein, \emph{{Einheitliche Feldtheorie von Gravitation und
  Elektrizit{\"a}t}. ({German}) [{Unified Field Theory} of gravitation and
  electricity]}, {\emph{Sitzungber. Preuss. Akad. Wiss.} {\bfseries 22} (1925)
  414}.

\bibitem{Palatini}
A.~Palatini, \emph{Deduzione invariantiva delle equazioni gravitazionali dal
  principio di hamilton},
  \href{https://doi.org/10.1007/BF03014670}{\emph{Rendiconti del Circolo
  Matematico di Palermo (1884-1940)} {\bfseries 43} (1919) 203}.

\bibitem{Tenkanen:2020dge}
T.~Tenkanen, \emph{{Tracing the high energy theory of gravity: an introduction
  to Palatini inflation}},
  \href{https://doi.org/10.1007/s10714-020-02682-2}{\emph{Gen. Rel. Grav.}
  {\bfseries 52} (2020) 33} [\href{https://arxiv.org/abs/2001.10135}{{\ttfamily
  2001.10135}}].

\bibitem{Bauer:2008zj}
F.~Bauer and D.A.~Demir, \emph{{Inflation with Non-Minimal Coupling: Metric
  versus Palatini Formulations}},
  \href{https://doi.org/10.1016/j.physletb.2008.06.014}{\emph{Phys. Lett. B}
  {\bfseries 665} (2008) 222}
  [\href{https://arxiv.org/abs/0803.2664}{{\ttfamily 0803.2664}}].

\bibitem{Planck:2018jri}
{\scshape Planck} collaboration, \emph{{Planck 2018 results. X. Constraints on
  inflation}}, \href{https://doi.org/10.1051/0004-6361/201833887}{\emph{Astron.
  Astrophys.} {\bfseries 641} (2020) A10}
  [\href{https://arxiv.org/abs/1807.06211}{{\ttfamily 1807.06211}}].

\bibitem{Rubio:2018ogq}
J.~Rubio, \emph{{Higgs inflation}},
  \href{https://doi.org/10.3389/fspas.2018.00050}{\emph{Front. Astron. Space
  Sci.} {\bfseries 5} (2019) 50}
  [\href{https://arxiv.org/abs/1807.02376}{{\ttfamily 1807.02376}}].

\bibitem{Burgess:2009ea}
C.P.~Burgess, H.M.~Lee and M.~Trott, \emph{{Power-counting and the Validity of
  the Classical Approximation During Inflation}},
  \href{https://doi.org/10.1088/1126-6708/2009/09/103}{\emph{JHEP} {\bfseries
  09} (2009) 103} [\href{https://arxiv.org/abs/0902.4465}{{\ttfamily
  0902.4465}}].

\bibitem{Barbon:2009ya}
J.L.F.~Barbon and J.R.~Espinosa, \emph{{On the Naturalness of Higgs
  Inflation}}, \href{https://doi.org/10.1103/PhysRevD.79.081302}{\emph{Phys.
  Rev. D} {\bfseries 79} (2009) 081302}
  [\href{https://arxiv.org/abs/0903.0355}{{\ttfamily 0903.0355}}].

\bibitem{Barvinsky:2009ii}
A.O.~Barvinsky, A.Y.~Kamenshchik, C.~Kiefer, A.A.~Starobinsky and
  C.F.~Steinwachs, \emph{{Higgs boson, renormalization group, and naturalness
  in cosmology}},
  \href{https://doi.org/10.1140/epjc/s10052-012-2219-3}{\emph{Eur. Phys. J. C}
  {\bfseries 72} (2012) 2219}
  [\href{https://arxiv.org/abs/0910.1041}{{\ttfamily 0910.1041}}].

\bibitem{Burgess:2010zq}
C.P.~Burgess, H.M.~Lee and M.~Trott, \emph{{Comment on Higgs Inflation and
  Naturalness}}, \href{https://doi.org/10.1007/JHEP07(2010)007}{\emph{JHEP}
  {\bfseries 07} (2010) 007} [\href{https://arxiv.org/abs/1002.2730}{{\ttfamily
  1002.2730}}].

\bibitem{Hertzberg:2010dc}
M.P.~Hertzberg, \emph{{On Inflation with Non-minimal Coupling}},
  \href{https://doi.org/10.1007/JHEP11(2010)023}{\emph{JHEP} {\bfseries 11}
  (2010) 023} [\href{https://arxiv.org/abs/1002.2995}{{\ttfamily 1002.2995}}].

\bibitem{Bezrukov:2010jz}
F.~Bezrukov, A.~Magnin, M.~Shaposhnikov and S.~Sibiryakov, \emph{{Higgs
  inflation: consistency and generalisations}},
  \href{https://doi.org/10.1007/JHEP01(2011)016}{\emph{JHEP} {\bfseries 01}
  (2011) 016} [\href{https://arxiv.org/abs/1008.5157}{{\ttfamily 1008.5157}}].

\bibitem{Bauer:2010jg}
F.~Bauer and D.A.~Demir, \emph{{Higgs-Palatini Inflation and Unitarity}},
  \href{https://doi.org/10.1016/j.physletb.2011.03.042}{\emph{Phys. Lett. B}
  {\bfseries 698} (2011) 425}
  [\href{https://arxiv.org/abs/1012.2900}{{\ttfamily 1012.2900}}].

\bibitem{Lerner:2011it}
R.N.~Lerner and J.~McDonald, \emph{{Unitarity-Violation in Generalized Higgs
  Inflation Models}},
  \href{https://doi.org/10.1088/1475-7516/2012/11/019}{\emph{JCAP} {\bfseries
  11} (2012) 019} [\href{https://arxiv.org/abs/1112.0954}{{\ttfamily
  1112.0954}}].

\bibitem{Kehagias:2013mya}
A.~Kehagias, A.~Moradinezhad~Dizgah and A.~Riotto, \emph{{Remarks on the
  Starobinsky model of inflation and its descendants}},
  \href{https://doi.org/10.1103/PhysRevD.89.043527}{\emph{Phys. Rev. D}
  {\bfseries 89} (2014) 043527}
  [\href{https://arxiv.org/abs/1312.1155}{{\ttfamily 1312.1155}}].

\bibitem{Ren:2014sya}
J.~Ren, Z.-Z.~Xianyu and H.-J.~He, \emph{{Higgs Gravitational Interaction, Weak
  Boson Scattering, and Higgs Inflation in Jordan and Einstein Frames}},
  \href{https://doi.org/10.1088/1475-7516/2014/06/032}{\emph{JCAP} {\bfseries
  06} (2014) 032} [\href{https://arxiv.org/abs/1404.4627}{{\ttfamily
  1404.4627}}].

\bibitem{McDonald:2020lpz}
J.~McDonald, \emph{{Does Palatini Higgs Inflation Conserve Unitarity?}},
  \href{https://doi.org/10.1088/1475-7516/2021/04/069}{\emph{JCAP} {\bfseries
  04} (2021) 069} [\href{https://arxiv.org/abs/2007.04111}{{\ttfamily
  2007.04111}}].

\bibitem{Hamada:2020kuy}
Y.~Hamada, K.~Kawana and A.~Scherlis, \emph{{On Preheating in Higgs
  Inflation}}, \href{https://doi.org/10.1088/1475-7516/2021/03/062}{\emph{JCAP}
  {\bfseries 03} (2021) 062}
  [\href{https://arxiv.org/abs/2007.04701}{{\ttfamily 2007.04701}}].

\bibitem{Antoniadis:2021axu}
I.~Antoniadis, A.~Guillen and K.~Tamvakis, \emph{{Ultraviolet behaviour of
  Higgs inflation models}},
  \href{https://doi.org/10.1007/JHEP08(2021)018}{\emph{JHEP} {\bfseries 08}
  (2021) 018} [\href{https://arxiv.org/abs/2106.09390}{{\ttfamily
  2106.09390}}].

\bibitem{Ema:2016dny}
Y.~Ema, R.~Jinno, K.~Mukaida and K.~Nakayama, \emph{{Violent Preheating in
  Inflation with Nonminimal Coupling}},
  \href{https://doi.org/10.1088/1475-7516/2017/02/045}{\emph{JCAP} {\bfseries
  02} (2017) 045} [\href{https://arxiv.org/abs/1609.05209}{{\ttfamily
  1609.05209}}].

\bibitem{Giudice:2010ka}
G.F.~Giudice and H.M.~Lee, \emph{{Unitarizing Higgs Inflation}},
  \href{https://doi.org/10.1016/j.physletb.2010.10.035}{\emph{Phys. Lett. B}
  {\bfseries 694} (2011) 294}
  [\href{https://arxiv.org/abs/1010.1417}{{\ttfamily 1010.1417}}].

\bibitem{Salvio:2015kka}
A.~Salvio and A.~Mazumdar, \emph{{Classical and Quantum Initial Conditions for
  Higgs Inflation}},
  \href{https://doi.org/10.1016/j.physletb.2015.09.020}{\emph{Phys. Lett. B}
  {\bfseries 750} (2015) 194}
  [\href{https://arxiv.org/abs/1506.07520}{{\ttfamily 1506.07520}}].

\bibitem{Ema:2017rqn}
Y.~Ema, \emph{{Higgs Scalaron Mixed Inflation}},
  \href{https://doi.org/10.1016/j.physletb.2017.04.060}{\emph{Phys. Lett. B}
  {\bfseries 770} (2017) 403}
  [\href{https://arxiv.org/abs/1701.07665}{{\ttfamily 1701.07665}}].

\bibitem{Gorbunov:2018llf}
D.~Gorbunov and A.~Tokareva, \emph{{Scalaron the healer: removing the
  strong-coupling in the Higgs- and Higgs-dilaton inflations}},
  \href{https://doi.org/10.1016/j.physletb.2018.11.015}{\emph{Phys. Lett. B}
  {\bfseries 788} (2019) 37}
  [\href{https://arxiv.org/abs/1807.02392}{{\ttfamily 1807.02392}}].

\bibitem{Ema:2020zvg}
Y.~Ema, K.~Mukaida and J.~van~de Vis, \emph{{Higgs inflation as nonlinear sigma
  model and scalaron as its $\sigma$-meson}},
  \href{https://doi.org/10.1007/JHEP11(2020)011}{\emph{JHEP} {\bfseries 11}
  (2020) 011} [\href{https://arxiv.org/abs/2002.11739}{{\ttfamily
  2002.11739}}].

\bibitem{Jinno:2019und}
R.~Jinno, M.~Kubota, K.-y.~Oda and S.C.~Park, \emph{{Higgs inflation in metric
  and Palatini formalisms: Required suppression of higher dimensional
  operators}}, \href{https://doi.org/10.1088/1475-7516/2020/03/063}{\emph{JCAP}
  {\bfseries 03} (2020) 063}
  [\href{https://arxiv.org/abs/1904.05699}{{\ttfamily 1904.05699}}].

\bibitem{Takahashi:2018brt}
T.~Takahashi and T.~Tenkanen, \emph{{Towards distinguishing variants of
  non-minimal inflation}},
  \href{https://doi.org/10.1088/1475-7516/2019/04/035}{\emph{JCAP} {\bfseries
  04} (2019) 035} [\href{https://arxiv.org/abs/1812.08492}{{\ttfamily
  1812.08492}}].

\bibitem{Hazumi:2019lys}
M.~Hazumi et~al., \emph{{LiteBIRD: A Satellite for the Studies of B-Mode
  Polarization and Inflation from Cosmic Background Radiation Detection}},
  \href{https://doi.org/10.1007/s10909-019-02150-5}{\emph{J. Low Temp. Phys.}
  {\bfseries 194} (2019) 443}.

\bibitem{CMB-S4:2016ple}
{\scshape CMB-S4} collaboration, \emph{{CMB-S4 Science Book, First Edition}},
  \href{https://arxiv.org/abs/1610.02743}{{\ttfamily 1610.02743}}.

\bibitem{CMB-S4:2020lpa}
{\scshape CMB-S4} collaboration, \emph{{CMB-S4: Forecasting Constraints on
  Primordial Gravitational Waves}},
  \href{https://arxiv.org/abs/2008.12619}{{\ttfamily 2008.12619}}.

\bibitem{Lyth:1998xn}
D.H.~Lyth and A.~Riotto, \emph{{Particle physics models of inflation and the
  cosmological density perturbation}},
  \href{https://doi.org/10.1016/S0370-1573(98)00128-8}{\emph{Phys. Rept.}
  {\bfseries 314} (1999) 1}
  [\href{https://arxiv.org/abs/hep-ph/9807278}{{\ttfamily hep-ph/9807278}}].

\bibitem{Planck:2018vyg}
{\scshape Planck} collaboration, \emph{{Planck 2018 results. VI. Cosmological
  parameters}},
  \href{https://doi.org/10.1051/0004-6361/201833910}{\emph{Astron. Astrophys.}
  {\bfseries 641} (2020) A6}
  [\href{https://arxiv.org/abs/1807.06209}{{\ttfamily 1807.06209}}].

\bibitem{Hamada:2014iga}
Y.~Hamada, H.~Kawai, K.-y.~Oda and S.C.~Park, \emph{{Higgs Inflation is Still
  Alive after the Results from BICEP2}},
  \href{https://doi.org/10.1103/PhysRevLett.112.241301}{\emph{Phys. Rev. Lett.}
  {\bfseries 112} (2014) 241301}
  [\href{https://arxiv.org/abs/1403.5043}{{\ttfamily 1403.5043}}].

\bibitem{Bezrukov:2014bra}
F.~Bezrukov and M.~Shaposhnikov, \emph{{Higgs inflation at the critical
  point}}, \href{https://doi.org/10.1016/j.physletb.2014.05.074}{\emph{Phys.
  Lett. B} {\bfseries 734} (2014) 249}
  [\href{https://arxiv.org/abs/1403.6078}{{\ttfamily 1403.6078}}].

\bibitem{Hamada:2014wna}
Y.~Hamada, H.~Kawai, K.-y.~Oda and S.C.~Park, \emph{{Higgs inflation from
  Standard Model criticality}},
  \href{https://doi.org/10.1103/PhysRevD.91.053008}{\emph{Phys. Rev. D}
  {\bfseries 91} (2015) 053008}
  [\href{https://arxiv.org/abs/1408.4864}{{\ttfamily 1408.4864}}].

\bibitem{Enckell:2020lvn}
V.-M.~Enckell, S.~Nurmi, S.~R\"as\"anen and E.~Tomberg, \emph{{Critical point
  Higgs inflation in the Palatini formulation}},
  \href{https://doi.org/10.1007/JHEP04(2021)059}{\emph{JHEP} {\bfseries 04}
  (2021) 059} [\href{https://arxiv.org/abs/2012.03660}{{\ttfamily
  2012.03660}}].

\bibitem{DeCross:2016cbs}
M.P.~DeCross, D.I.~Kaiser, A.~Prabhu, C.~Prescod-Weinstein and
  E.I.~Sfakianakis, \emph{{Preheating after multifield inflation with
  nonminimal couplings, III: Dynamical spacetime results}},
  \href{https://doi.org/10.1103/PhysRevD.97.023528}{\emph{Phys. Rev. D}
  {\bfseries 97} (2018) 023528}
  [\href{https://arxiv.org/abs/1610.08916}{{\ttfamily 1610.08916}}].

\bibitem{Sfakianakis:2018lzf}
E.I.~Sfakianakis and J.~van~de Vis, \emph{{Preheating after Higgs Inflation:
  Self-Resonance and Gauge boson production}},
  \href{https://doi.org/10.1103/PhysRevD.99.083519}{\emph{Phys. Rev. D}
  {\bfseries 99} (2019) 083519}
  [\href{https://arxiv.org/abs/1810.01304}{{\ttfamily 1810.01304}}].

\bibitem{Alonso:2015fsp}
R.~Alonso, E.E.~Jenkins and A.V.~Manohar, \emph{{A Geometric Formulation of
  Higgs Effective Field Theory: Measuring the Curvature of Scalar Field
  Space}}, \href{https://doi.org/10.1016/j.physletb.2016.01.041}{\emph{Phys.
  Lett. B} {\bfseries 754} (2016) 335}
  [\href{https://arxiv.org/abs/1511.00724}{{\ttfamily 1511.00724}}].

\bibitem{Nagai:2019tgi}
R.~Nagai, M.~Tanabashi, K.~Tsumura and Y.~Uchida, \emph{{Symmetry and geometry
  in a generalized Higgs effective field theory: Finiteness of oblique
  corrections versus perturbative unitarity}},
  \href{https://doi.org/10.1103/PhysRevD.100.075020}{\emph{Phys. Rev. D}
  {\bfseries 100} (2019) 075020}
  [\href{https://arxiv.org/abs/1904.07618}{{\ttfamily 1904.07618}}].

\bibitem{Ema:2021xhq}
Y.~Ema, R.~Jinno, K.~Nakayama and J.~van~de Vis, \emph{{Preheating from target
  space curvature and unitarity violation: Analysis in field space}},
  \href{https://doi.org/10.1103/PhysRevD.103.103536}{\emph{Phys. Rev. D}
  {\bfseries 103} (2021) 103536}
  [\href{https://arxiv.org/abs/2102.12501}{{\ttfamily 2102.12501}}].

\bibitem{Lerner:2010mq}
R.N.~Lerner and J.~McDonald, \emph{{A Unitarity-Conserving Higgs Inflation
  Model}}, \href{https://doi.org/10.1103/PhysRevD.82.103525}{\emph{Phys. Rev.
  D} {\bfseries 82} (2010) 103525}
  [\href{https://arxiv.org/abs/1005.2978}{{\ttfamily 1005.2978}}].

\bibitem{Barbon:2015fla}
J.L.F.~Barbon, J.A.~Casas, J.~Elias-Miro and J.R.~Espinosa, \emph{{Higgs
  Inflation as a Mirage}},
  \href{https://doi.org/10.1007/JHEP09(2015)027}{\emph{JHEP} {\bfseries 09}
  (2015) 027} [\href{https://arxiv.org/abs/1501.02231}{{\ttfamily
  1501.02231}}].

\bibitem{Lee:2018esk}
H.M.~Lee, \emph{{Light inflaton completing Higgs inflation}},
  \href{https://doi.org/10.1103/PhysRevD.98.015020}{\emph{Phys. Rev. D}
  {\bfseries 98} (2018) 015020}
  [\href{https://arxiv.org/abs/1802.06174}{{\ttfamily 1802.06174}}].

\bibitem{Lee:2021dgi}
H.M.~Lee and A.G.~Menkara, \emph{{Cosmology of linear Higgs-sigma models with
  conformal invariance}},
  \href{https://doi.org/10.1007/JHEP09(2021)018}{\emph{JHEP} {\bfseries 09}
  (2021) 018} [\href{https://arxiv.org/abs/2104.10390}{{\ttfamily
  2104.10390}}].

\bibitem{tHooft:1974toh}
G.~'t~Hooft and M.J.G.~Veltman, \emph{{One loop divergencies in the theory of
  gravitation}}, {\emph{Ann. Inst. H. Poincare Phys. Theor. A} {\bfseries 20}
  (1974) 69}.

\bibitem{Donoghue:1995cz}
J.F.~Donoghue, \emph{{Introduction to the effective field theory description of
  gravity}},  in \emph{{Advanced School on Effective Theories}}, 6, 1995
  [\href{https://arxiv.org/abs/gr-qc/9512024}{{\ttfamily gr-qc/9512024}}].

\bibitem{Kannike:2015apa}
K.~Kannike, G.~H\"utsi, L.~Pizza, A.~Racioppi, M.~Raidal, A.~Salvio et~al.,
  \emph{{Dynamically Induced Planck Scale and Inflation}},
  \href{https://doi.org/10.1007/JHEP05(2015)065}{\emph{JHEP} {\bfseries 05}
  (2015) 065} [\href{https://arxiv.org/abs/1502.01334}{{\ttfamily
  1502.01334}}].

\bibitem{Salvio:2016vxi}
A.~Salvio, \emph{{Solving the Standard Model Problems in Softened Gravity}},
  \href{https://doi.org/10.1103/PhysRevD.94.096007}{\emph{Phys. Rev. D}
  {\bfseries 94} (2016) 096007}
  [\href{https://arxiv.org/abs/1608.01194}{{\ttfamily 1608.01194}}].

\bibitem{He:2018gyf}
M.~He, A.A.~Starobinsky and J.~Yokoyama, \emph{{Inflation in the mixed
  Higgs-$R^2$ model}},
  \href{https://doi.org/10.1088/1475-7516/2018/05/064}{\emph{JCAP} {\bfseries
  05} (2018) 064} [\href{https://arxiv.org/abs/1804.00409}{{\ttfamily
  1804.00409}}].

\bibitem{He:2018mgb}
M.~He, R.~Jinno, K.~Kamada, S.C.~Park, A.A.~Starobinsky and J.~Yokoyama,
  \emph{{On the violent preheating in the mixed Higgs-$R^2$ inflationary
  model}}, \href{https://doi.org/10.1016/j.physletb.2019.02.008}{\emph{Phys.
  Lett. B} {\bfseries 791} (2019) 36}
  [\href{https://arxiv.org/abs/1812.10099}{{\ttfamily 1812.10099}}].

\bibitem{He:2020ivk}
M.~He, R.~Jinno, K.~Kamada, A.A.~Starobinsky and J.~Yokoyama, \emph{{Occurrence
  of tachyonic preheating in the mixed Higgs-R$^2$ model}},
  \href{https://doi.org/10.1088/1475-7516/2021/01/066}{\emph{JCAP} {\bfseries
  01} (2021) 066} [\href{https://arxiv.org/abs/2007.10369}{{\ttfamily
  2007.10369}}].

\bibitem{He:2020qcb}
M.~He, \emph{{Perturbative Reheating in the Mixed Higgs-$R^2$ Model}},
  \href{https://doi.org/10.1088/1475-7516/2021/05/021}{\emph{JCAP} {\bfseries
  05} (2021) 021} [\href{https://arxiv.org/abs/2010.11717}{{\ttfamily
  2010.11717}}].

\bibitem{Gialamas:2019nly}
I.D.~Gialamas and A.B.~Lahanas, \emph{{Reheating in $R^2$ Palatini inflationary
  models}}, \href{https://doi.org/10.1103/PhysRevD.101.084007}{\emph{Phys. Rev.
  D} {\bfseries 101} (2020) 084007}
  [\href{https://arxiv.org/abs/1911.11513}{{\ttfamily 1911.11513}}].

\bibitem{Gialamas:2020vto}
I.D.~Gialamas, A.~Karam, A.~Lykkas and T.D.~Pappas, \emph{{Palatini-Higgs
  inflation with nonminimal derivative coupling}},
  \href{https://doi.org/10.1103/PhysRevD.102.063522}{\emph{Phys. Rev. D}
  {\bfseries 102} (2020) 063522}
  [\href{https://arxiv.org/abs/2008.06371}{{\ttfamily 2008.06371}}].

\bibitem{Gialamas:2021enw}
I.D.~Gialamas, A.~Karam, T.D.~Pappas and V.C.~Spanos, \emph{{Scale-invariant
  quadratic gravity and inflation in the Palatini formalism}},
  \href{https://doi.org/10.1103/PhysRevD.104.023521}{\emph{Phys. Rev. D}
  {\bfseries 104} (2021) 023521}
  [\href{https://arxiv.org/abs/2104.04550}{{\ttfamily 2104.04550}}].

\bibitem{DeFelice:2010aj}
A.~De~Felice and S.~Tsujikawa, \emph{{f(R) theories}},
  \href{https://doi.org/10.12942/lrr-2010-3}{\emph{Living Rev. Rel.} {\bfseries
  13} (2010) 3} [\href{https://arxiv.org/abs/1002.4928}{{\ttfamily
  1002.4928}}].

\bibitem{Edery:2019txq}
A.~Edery and Y.~Nakayama, \emph{{Palatini formulation of pure $R^2$ gravity
  yields Einstein gravity with no massless scalar}},
  \href{https://doi.org/10.1103/PhysRevD.99.124018}{\emph{Phys. Rev. D}
  {\bfseries 99} (2019) 124018}
  [\href{https://arxiv.org/abs/1902.07876}{{\ttfamily 1902.07876}}].

\bibitem{Antoniadis:2018ywb}
I.~Antoniadis, A.~Karam, A.~Lykkas and K.~Tamvakis, \emph{{Palatini inflation
  in models with an $R^2$ term}},
  \href{https://doi.org/10.1088/1475-7516/2018/11/028}{\emph{JCAP} {\bfseries
  11} (2018) 028} [\href{https://arxiv.org/abs/1810.10418}{{\ttfamily
  1810.10418}}].

\bibitem{Tenkanen:2019jiq}
T.~Tenkanen, \emph{{Minimal Higgs inflation with an $R^2$ term in Palatini
  gravity}}, \href{https://doi.org/10.1103/PhysRevD.99.063528}{\emph{Phys. Rev.
  D} {\bfseries 99} (2019) 063528}
  [\href{https://arxiv.org/abs/1901.01794}{{\ttfamily 1901.01794}}].

\bibitem{Lykkas:2021vax}
A.~Lykkas and K.~Tamvakis, \emph{{Extended interactions in the Palatini-$R^2$
  inflation}}, \href{https://doi.org/10.1088/1475-7516/2021/08/043}{\emph{JCAP}
  {\bfseries 08} (2021) } [\href{https://arxiv.org/abs/2103.10136}{{\ttfamily
  2103.10136}}].

\bibitem{Kleyn:2004yj}
A.~Kleyn, \emph{{Metric-affine manifold}},
  \href{https://arxiv.org/abs/gr-qc/0405028}{{\ttfamily gr-qc/0405028}}.

\bibitem{Iosifidis:2018zwo}
D.~Iosifidis and T.~Koivisto, \emph{{Scale transformations in metric-affine
  geometry}}, \href{https://doi.org/10.3390/universe5030082}{\emph{Universe}
  {\bfseries 5} (2019) 82} [\href{https://arxiv.org/abs/1810.12276}{{\ttfamily
  1810.12276}}].

\end{thebibliography}\endgroup
\end{document}